\documentclass{aastex62}

\received{July 9, 2018}
\revised{October 9, 2018}
\accepted{\today}
\submitjournal{ApJS}

\shorttitle{An empirical template library }
\shortauthors{Du et al.}

\begin{document}

\title{An Empirical Template Library for FGK and Late--type A Stars Using LAMOST Observed Spectra}

\correspondingauthor{A-Li Luo}
\email{lal@bao.ac.cn}

\author[0000-0002-0786-7307]{Bing Du}
\affil{Key Laboratory of Optical Astronomy, National Astronomical Observatories, Chinese Academy of Sciences, Beijing 100012, China.}
\affil{University of Chinese Academy of Sciences, Beijing 100049, China.}

\author[0000-0001-7865-2648]{A-Li Luo}
\affiliation{Key Laboratory of Optical Astronomy, National Astronomical Observatories, Chinese Academy of Sciences, Beijing 100012, China.}
\affiliation{University of Chinese Academy of Sciences, Beijing 100049, China.}

\author{F. Zuo}
\affiliation{Key Laboratory of Optical Astronomy, National Astronomical Observatories, Chinese Academy of Sciences, Beijing 100012, China.}
\affiliation{University of Chinese Academy of Sciences, Beijing 100049, China.}

\author{Z-R. Bai}
\affiliation{Key Laboratory of Optical Astronomy, National Astronomical Observatories, Chinese Academy of Sciences, Beijing 100012, China.}

\author{R. Wang}
\affiliation{Key Laboratory of Optical Astronomy, National Astronomical Observatories, Chinese Academy of Sciences, Beijing 100012, China.}
\affiliation{University of Chinese Academy of Sciences, Beijing 100049, China.}

\author{Y-H. Song}
\affiliation{Key Laboratory of Optical Astronomy, National Astronomical Observatories, Chinese Academy of Sciences, Beijing 100012, China.}

\author{W. Hou}
\affiliation{Key Laboratory of Optical Astronomy, National Astronomical Observatories, Chinese Academy of Sciences, Beijing 100012, China.}

\author{Y-B. Li}
\affiliation{Key Laboratory of Optical Astronomy, National Astronomical Observatories, Chinese Academy of Sciences, Beijing 100012, China.}

\author{J-N. Zhang}
\affiliation{Key Laboratory of Optical Astronomy, National Astronomical Observatories, Chinese Academy of Sciences, Beijing 100012, China.}

\author{Y-X. Guo}
\affiliation{Key Laboratory of Optical Astronomy, National Astronomical Observatories, Chinese Academy of Sciences, Beijing 100012, China.}
\affiliation{University of Chinese Academy of Sciences, Beijing 100049, China.}

\author{J-J. Chen}
\affiliation{Key Laboratory of Optical Astronomy, National Astronomical Observatories, Chinese Academy of Sciences, Beijing 100012, China.}

\author{M-X. Wang}
\affiliation{Key Laboratory of Optical Astronomy, National Astronomical Observatories, Chinese Academy of Sciences, Beijing 100012, China.}
\affiliation{University of Chinese Academy of Sciences, Beijing 100049, China.}
\author{Y-F. Wang}
\affiliation{Key Laboratory of Optical Astronomy, National Astronomical Observatories, Chinese Academy of Sciences, Beijing 100012, China.}
\author{X. Kong}
\affiliation{Key Laboratory of Optical Astronomy, National Astronomical Observatories, Chinese Academy of Sciences, Beijing 100012, China.}
\affiliation{University of Chinese Academy of Sciences, Beijing 100049, China.}

\author{K-F. Wu}
\affiliation{Key Laboratory of Optical Astronomy, National Astronomical Observatories, Chinese Academy of Sciences, Beijing 100012, China.}
\affiliation{University of Chinese Academy of Sciences, Beijing 100049, China.}

\author{X. Wang}
\affiliation{Key Laboratory of Optical Astronomy, National Astronomical Observatories, Chinese Academy of Sciences, Beijing 100012, China.}

\author{Y. Wu}
\affiliation{Key Laboratory of Optical Astronomy, National Astronomical Observatories, Chinese Academy of Sciences, Beijing 100012, China.}

\author{Y.-H Hou}
\affiliation{Nanjing Institute of Astronomical Optics \& Technology, National Astronomical Observatories, Chinese Academy of Sciences, Nanjing 210042, China}

\author{Y-H. Zhao}
\affiliation{Key Laboratory of Optical Astronomy, National Astronomical Observatories, Chinese Academy of Sciences, Beijing 100012, China.}



\begin{abstract}

We present an empirical stellar spectra library created using spectra from the Large Sky Area Multi-Object Fiber Spectroscopic Telescope (LAMOST)  DR5. This library represents a uniform data set ranging from 3750 through 8500~K in effective temperature ($T_{\rm eff}$), from -2.5 through +1.0 dex in metallicity ([Fe/H]), and from 0 to 5.0 dex in gravity (log \emph{g}). The spectra in the library have resolution R $\sim$1800, with well--calibrated fluxes and rest--framed wavelengths. Using a large number of red stars observed by LAMOST, we generated denser K type templates to fill in data missing from current empirical spectral libraries, particularly the late K type. For K giants, we calibrated the spectroscopic surface gravities against the asteroseismic surface gravities. To verify the reliability of the parameters labeled for this library, we performed an internal cross-validation by using a $\chi^2$ minimization method to interpolate parameters of each individual spectrum using the remaining spectra in the library. We obtained precisions of 41 K, 0.11 dex, and 0.05 dex for $T_{\rm eff}$, log \emph{g}, and [Fe/H],  respectively, which means the templates are labeled with correct stellar parameters. Through external comparisons, we confirmed that measurements of  the stellar parameters through this library can achieve accuracies of approximately 125K in $T_{\rm eff}$, 0.1 dex in [Fe/H] and 0.20 dex in log \emph{g} without systematic offset. This empirical library is useful for stellar parameter measurements because it has large parameter coverage and full wavelength coverage from 3800 to 8900 \AA.
\end{abstract}

\keywords{techniques, spectroscopic -- methods, data analysis--methods: statistical}


\section{Introduction} \label{sec:intro}

Measuring stellar properties such as effective temperature, metallicity, and surface gravity,  is an important and long--standing problem in astronomy.  The masses, surface gravities, and temperatures of stars are benchmarks against which we test models of stellar structure and evolution. The abundances of iron and other elements in stellar populations help trace the nucleosynthetic enrichment history of the Milky Way \citep{yee}. However, the determination of stellar properties is often completed by comparisons to empirical templates with known parameters, or synthetic spectra.

Synthetic template libraries are widely used in measurements of stellar parameters including the effective temperature ($T_{\rm eff}$), the logarithm of the surface gravity (log \emph{g}) and the metallicity ([Fe/H]),  due to their wide span of parameter space \citep{Bailer, Willemsen, lee, aspcap}. Many of these complete synthetic libraries are available to the public \citep{kurucz, meszaros, husser}. However,  the models used to create these synthetic spectra are based on basic assumptions, such as local thermal equilibrium, plane parallel, and pressure broadening models etc., which somewhat differ from the actual physical environments of stars. In addition, incomplete lists of line opacities may also affect the synthetic spectra,  particularly for cool stars, which have complex, molecular-rich atmospheres. All of the aforementioned factors will lead to inconsistency between the model and observed spectra at the wavelengths of some features. Therefore, empirical libraries are still extremely important for both constraining models and measuring stellar parameters where the models are limited.

Empirical spectral libraries obtained through observations of real stars  have a rich history in stellar astronomy and have important applications in different fields. They are used as references to classify stars and to determine atmospheric parameters \citep{katz,prugniel,koleva,wua,wub,xiang,kesseli, yee}. They are also important ingredients in the  modeling of stellar populations, which are used to study the history of galaxies \citep{leborgne, prugniel07}.  The most important characteristics of an empirical library are (1) the distribution of the stars in the parameter space with axes of $T_{\rm eff}$, log \emph{g} and [Fe/H]; (2) the wavelength coverage; (3) the spectral resolution. Other properties are also to be considered, such as the precision and uniformity of the wavelength calibration and spectral resolution, and the accuracy of the flux calibration. Existing empirical spectral libraries are more limited  than theoretical libraries in their coverage of stellar parameter space.  Moreover, the spectral resolution and wavelength coverage are also limited {owing to the instrument capability. Some empirical libraries include only certain stellar types \citep{hawley,coaddmethod,schmidt}.  Other libraries such as ELODIE \citep{prugniel}, UVES-POP \citep{uvespop}, INDO-US \citep{cflib}, MILES \citep{miles}, Yee \citep{yee}  and  Kesseli \citep{kesseli},  aim to cover a wide range of stellar types. However, all of these libraries have shortcomings. Their specifications are summarized in Table \ref{tab_libs}.

\begin{table}
\begin{center}
\caption{Summary of existing optical empirical libraries}\label{tab_libs}
 \begin{tabular}{lllllll}
 
\hline\noalign{\smallskip} \hline\noalign{\smallskip}
 Empirical  libraries &   Number of stars  &  R  &  Wavelength coverage (\AA)  &  Flux calibration  &  Poor parameter coverage \\
\hline\noalign{\smallskip}
ELODIE & 1388  &  10000  &  3900--6800 &  Good   &  Giants \& Late-K Dwarfs   \\ 
UVES-POP & 400  &  80,000 & 3000--10000 &  Good   &  Giants \& K Dwarfs\\
INDO-US  & 1273  & 5000 & 3460--9460 &  Poor  &  K dwarfs \\
MILES &  985   &  2000 & 3520--7500 &  Good  &  Giants \& K Dwarfs  \\
Yee et al. (2017) & 404 &  60,000 &  4990--6410 &  Rectified spectra &  [Fe/H] $< $ -0.5 \\
Kesseli et al. (2017)  & 324  & 2000  & 3650--10,200 &  Good  &  \textbf{Spectral \& luminosity} \\
 &  &   &  &    & \textbf{classes with metallicities} \\
 &  &   &  &    & no  $T_{\rm eff}$, no log \emph{g}\\
\noalign{\smallskip}\hline
\end{tabular}
\end{center}
\end{table}

The greatest limitation of empirical libraries  is their poor  parameter coverage. Limited by the observational conditions, we naturally get the type of stars and abundance patterns available within the solar neighborhood. The parameter coverage is quite incomplete. Improvement is needed, particularly regarding cool and low-metallicity stars. The advent of the  Large Sky Area Multi-Object Fiber Spectroscopic Telescope (LAMOST) enabled sampling of a larger area of the Milky Way  to further extend the parameter space of empirical  libraries. 

The last version of the  ELODIE library,  version 3.2, has been used by the LAMOST stellar parameter pipeline (LASP) to automatically determine the stellar parameters including $T_{\rm eff}$, log \emph{g}, [Fe/H], and radial velocity RV for FGK and late--type A stars \citep{lal,wub}. The ELODIE library  includes  1962 spectra of 1388 stars observed with the eponym Echelle spectrograph at a spectral resolution R $\sim$ 42, 000 ( $\Delta \lambda \sim 0.13 \AA $ ) in the wavelength range 3900 to 6800 \AA. The library is also available at a resolution of R $\sim$ 10, 000 ( $\Delta \lambda \sim 0.55 \AA $ ) for the population synthesis in  PegaseHR \citep{leborgne}. The temperature, gravity and metallicity (TGM) function is an interpolator of the ELODIE library, through which the interpolated template spectra ( R $\sim$ 10, 000 ) are  reconstructed. The interpolator consists of polynomial expansions of each wavelength element  in powers of log($T_{\rm eff}$), log \emph{g},  [Fe/H] and f($\sigma$) which is a function of the rotational broadening parameterized by $\sigma$ ( the standard deviation of a Gaussian) \citep{wub}. Where there is poor parameter coverage of giants and late-type K dwarfs, theoretical templates are added to build the interpolator of the ELODIE version 3.2 ( private communication with Philippe Prugniel ). To assess the accuracy of the ELODIE interpolator under LAMOST resolution ($R \sim 1800$ ), we performed an internal validation, where we treated each star in the library as an unknown target, the spectrum of which was degraded to the LAMOST resolution. We ran LASP to compute their parameters from the ELODIE interpolator and then compared these derived parameters to their library values. For the relatively incomplete K stars ($T_{\rm eff} \leq $ 5000 K), the differences between the derived and library values had a scatter of 140 K in $T_{\rm eff}$, 0.26 dex in log \emph{g},  and 0.13 dex in [Fe/H]. When restricting our analysis to the F and G stars (5000 K  $ <  T_{\rm eff} \leq $ 7500 K), the performance improved to  110 K in $T_{\rm eff}$, 0.18 dex in log \emph{g},  and 0.1 dex in [Fe/H]. This proves that some differences exist between the observed spectra and the theoretical models. The ELODIE library is also limited by its small wavelength coverage of 3900--6800 \AA,  which does not include the longer--wavelength features necessary for studying the  metallicities and surface gravities of cool stars ( e.g., Na I at 8200 \AA ).

For LAMOST DR5, LASP has determined  $T_{\rm eff}$, log \emph{g}, [Fe/H]  and RV for more than 5 million spectra,  the stellar parameters and spectra are both publically available\ \footnote{\url{dr5.lamost.org/}}. LASP employs two methods, Correlation Function Initial (CFI) guess \citep{du} and the ULySS: a full spectrum fitting package\ \footnote{\url{http://ulyss.univ-lyon1.fr/models.html/}} \citep{wub}. The initial coarse measurements of CFI were adopted as initial guesses for ULySS, which enabled ULySS to avoid converging to a local optimal solution.  LASP operates in two stages to measure the stellar parameters. In the first stage we estimated the stellar parameters from the original spectrum with continuum.  In the second stage we used} the rectified spectrum with the pseudo--continuum divided out. Because the quality of the pseudo--continuum used in the second stage depends on the credibility of the stellar parameters estimated in the first stage, the consistency between the results of the two stages indicates the reliability of the stellar parameters. The quoted accuracies were about 110 K, 0.25 dex,  and 0.11 dex  for $T_{\rm eff}$, log \emph{g}, and [Fe/H],  respectively,  in the effective temperature range of 3700 K  $\le  T_{\rm eff} \leq$ 8500 K  \citep{lal}. The systematic offset was negligible for all stars except for log \emph{g} of K giants ($T_{\rm eff} \leq $ 5000 K  and log \emph{g} $ \le $ 3.5 dex ). Because a large number of cool stars were observed by LAMOST,  we were able to generate  more K--type templates to overcome the lack of K type spectra in current empirical spectral libraries, particularly late K type.  For K giants, the spectroscopic surface gravities could be calibrated against the asteroseismic surface gravities,  and the offsets between them could be corrected. This empirical library has large parameter coverage including cool stars with $T_{\rm eff} <$ 4500 K  and full wavelength coverage from 3800 to 8900 \AA.  With this new stellar template library and its large parameter coverage, determinations of stellar parameters and spectral typing will improve. This set of templates can be used as reference stars to determine the stellar parameters and can also be used as a training set for more complicated machine learning spectral typing.  In addition, the templates can be used in stellar population synthesis models, which have long been used as a tool to assemble model galaxies from the co--addition of individual stellar spectra \citep{bruzual, prugniel07}.  

In this study, we present a library of empirical stellar spectra which are binned by effective temperature from 3750 K to 8500 K, metallicity from -2.5 dex to  +1.0 dex,  and log \emph{g} from 0 dex to 5.0 dex. The wavelength coverage for the templates is from 3800 to 8900 \AA \ at a resolution of R $\sim$ 1800. The advantage of this library is that it includes the most complete empirical spectra of K type stars, particularly K giants, the spectroscopic surface gravities of which were well calibrated against the asteroseismic surface gravities. To verify the reliability of the parameters labeled for this library,  we performed an internal cross--validation and a couple of external comparisons by using $\chi^2$ minimization in multidimensional parameter space. The results show that this library was labeled with accurate stellar parameters, with uncertainties of  41 K, 0.05 dex, and 0.11 dex for $T_{\rm eff}$, [Fe/H],  and log \emph{g}, respectively.  Moreover, this library represents a uniform standard dataset. The outline of this paper is as follows. In Section \ref{sec:library}, we detail the methods used for construction of the library. We describe the stellar parameters determined for each star type,  flux calibration and dereddening, the co--addition process,  and quality control. In Section \ref{sec:result}, we present the results of this library including the stellar parameter coverage, the template spectra,  and  individual spectral feature lines.  Next, we present the validations of the stellar parameters derived from this library. Finally, we draw conclusions for our templates in Section 5.

\section{Construction of the library} \label{sec:library}
\subsection{Data selection}
\subsubsection{F, G and Late--type A stars}

LASP adopts two methods, CFI and UlySS \citep{wub},  to determine the stellar parameters including $T_{\rm eff}$, log \emph{g}, [Fe/H] and RV. The CFI method provides initial guesses for ULySS to keep ULySS  from converging to a local optimal solution.  LASP produces reliable stellar parameters for F, G and late--type A stars \citep{lal}, with  reported global uncertainties of 110 K, 0.20 dex, and 0.10 dex  for $T_{\rm eff}$, log \emph{g}, and [Fe/H],  respectively,  in the effective temperature range of  5000 K $ \le  T_{\rm eff} \leq$ 8500 K  \citep{lal,gao}.  The LAMOST DR5 includes more than  4  million  F,  G, and late--type A stars. We selected stars with spectra having a signal-to-noise ratio (SNR) of the g band (SNRg) of at least 10.0 and with reported stellar parameters from LASP.  For late--type A and early--type F stars ($T_{\rm eff} > $ 6500 K), We excluded chemically peculiar stars  according to the characteristic of underabundance of calcium and overabundance of iron element of Am stars \citep{hou}.

\subsubsection{K Giants}

\begin{figure}
\centering
\includegraphics[width=16.0cm,height=16.0cm]{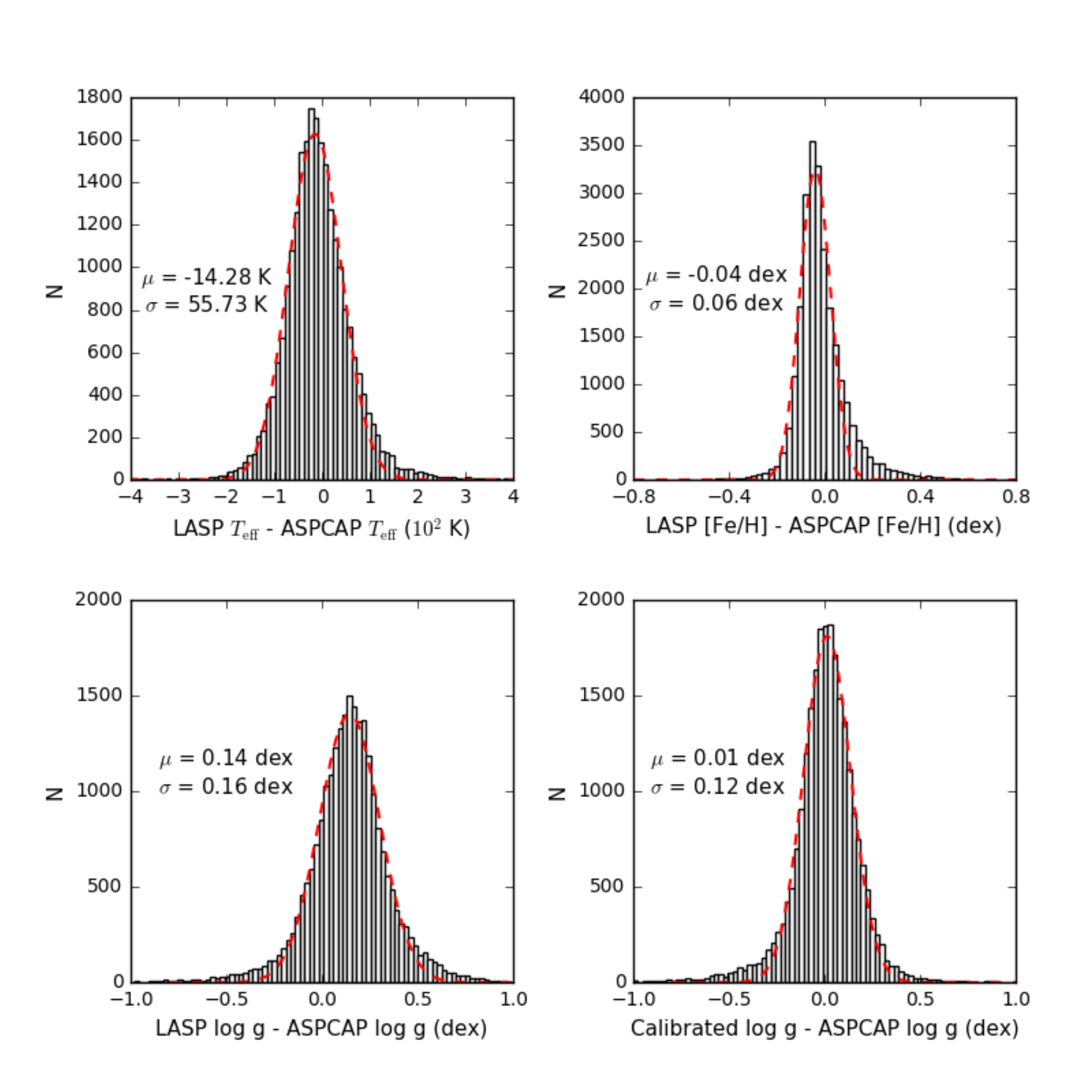}
\caption{ Histograms of differences between the LASP--derived parameters and the  ASPCAP--derived parameters from the 24,196 common targets. The red dashed curves are Gaussian fits to the distributions, and the mean and dispersion of the Gaussian fit to the mean and standard deviation values of the differences are also labeled. The lower right side shows the differences between the calibrated LASP--log \emph{g} and the  ASPCAP--log \emph{g}.}
\label{fig1}
\end{figure}

For K giants, we cross--matched LAMOST DR5 with the Apache Point Observatory Galactic Evolution Experiment (APOGEE) of the Sloan Digital Sky Survey (SDSS) DR14 because  APOGEE stars are predominantly giants.  The APOGEE Stellar Parameter and Chemical Abundances Pipeline (ASPCAP) determines atmospheric parameters and chemical abundances from observed spectra ( R $ \sim $ 22,500)  by comparing the observed spectra to libraries of theoretical spectra, using $\chi^2$  minimization in multidimensional parameter space. It provides  $T_{\rm eff}$, log \emph{g}, and [Fe/H] precise to 2\%, 0.1 dex, and 0.05 dex,  respectively aspcap. 
As shown in Fig \ref{fig1}, we compared the common targets of about 24, 196 spectra, including repeat observations,  that include both LASP measurements and ASPCAP results. In this comparison, the statistical fits characterize the quality of agreement. The effective temperatures  and  metallicities  of LASP  agreed with those of ASPCAP, to within 55 K and  0.1 dex,  respectively. However, a systematic difference was present between the gravities derived by LASP and ASPCAP such that the LASP results were systematically larger. The gravities of ASPCAP were calibrated against a sample of stars in the Kepler field with gravities from asteroseismic analysis. The spectroscopic surface gravities were systematically higher than the asteroseismic values at lower surface gravities \citep{holtzman}. The effect of such an existence of surface gravity offsets is currently not well understood, as no dependence on metallicity was found. Calibration of the LAMOST stellar surface gravities using the Kepler asteroseismic data was conducted through  detrending  for the clear trends between $\Delta$log \emph{g}( LAMOST-seismic) and spectroscopic  $T_{\rm eff}$,  as well as log \emph{g} \citep{wang}. We separated  red clump stars from red giant stars by using the relation $-0.0010 \times T_{\rm eff, LASP} + 7.10 < {\rm log}  \, \emph{g}_{\rm LASP} < -0.0005 \times T_{\rm eff, LASP} + 5.05 $ \citep{chen}. For red clump stars, the LAMOST stellar surface gravities agreed with  the asteroseismic surface gravities. For red giant stars, we calibrated the LASP--derived values of log \emph{g} by using the relation described in Wang et al. (2016).  The calibration relations  adopted from Chen et al. (2015) and Wang et al. (2016) are given below.

\begin{equation}
   {\rm log}\,g_{\rm corr} = {\rm log}\,g_{\rm LASP},   \;   \;  {\rm as} \; -0.0010 \times T_{\rm eff, LASP} + 7.10 < {\rm log} \emph{g}_{\rm LASP} < -0.0005 \times T_{\rm eff, LASP} + 5.05
 \label{eq:eq1}
\end{equation}
\begin{equation}
   {\rm log}\,g_{\rm corr} = {\rm log}\, g_{\rm LASP} - 5.716 + 1.283 \times  \frac{T_{\rm eff, LASP}}{1000} + 1.188 \times {\rm log}\, g_{\rm LASP} - 0.2882 \times \frac{T_{\rm eff, LASP}}{1000} \times  {\rm log}\, g_{\rm LASP}
\label{eq:eq2}
\end{equation}

The histogram at the lower right in Fig \ref{fig1} shows that the surface gravity offsets were negligible ($ \mu \sim$ 0.01 dex  $\ll \sigma \sim$ 0.12 dex) after calibrations using the relations given by Eq  \ref{eq:eq1} and Eq \ref{eq:eq2}. Therefore, we calibrated the LASP--derived log \emph{g} for K giants ( $T_{\rm eff} \le $ 5100 K, log \emph{g} $\le$ 3.5 dex), by using the  relations given by Eq \ref{eq:eq1} and Eq \ref{eq:eq2}. For the common targets that include both LASP measurements and ASPCAP results, the ASPCAP results were adopted. 

\subsubsection{K Dwarfs}
The K dwarf stellar parameters were consistent with those of high-resolution (HR) spectra (or Kepler asteroseismology, in the case of log \emph{g})  \citep{lal}. Therefore, the main consideration for K dwarfs is pollution caused by the surface gravity offsets of K giants. Firstly, we selected roughly K dwarfs from the LAMOST DR5 with  spectral SNRg $\geq$ 10.0  and log \emph{g} $ > 3.5 $ dex. We obtained 507, 505 spectra of K dwarfs, including repeat observations.  Secondly, we cross--matched the selected K dwarfs with the APOGEE giants (log \emph{g} $ < 3.5 $ dex), and also with K giants determined by using the method described in Liu et al. (2015). We abandoned the K dwarfs that were classified as K giants in the other two catalogs. Finally, we removed 101 spectra  from the 507, 505 spectra of K dwarfs. The proportion of the K giants pollution was negligible for large sample statistics.

\subsection{Back to Rest Frames}
 For  F, G,  K and late--type A stars, the spectroscopic radial velocities from LASP were used to shift all of the spectra into their rest frames. The quoted precision of the LASP radial velocities was about 4--6 ${\rm km \, s}^{-1}$ \citep{lal,gao}. The wavelength range of LAMOST covers 3700 to 9000 \AA \ and is recorded in two arms:  a blue arm (3700--5900 \AA) and a red arm (5700--9000 \AA), with a resolving power of R $\sim$ 1800.  A final spectrum was obtained by co--adding several exposures  and connecting wavelength bands.   On the original  charge--coupled device (CCD) image, 4 K pixels were used for recording blue or red wavelength regions, which means each \AA \  was sampled in two CCD pixels in the raw data. The final spectra were re--sampled in constant--velocity pixels, with a pixel scale of 69 ${\rm km \, s}^{-1}$ \citep{lal}. It should be noted that the precision of the LASP RV was about 4--6 ${\rm km \, s}^{-1}$, which means our RV calculations are accurate at sub--pixel values ($<$ 10\% of the pixel scale).  Therefore, we were able to obtain an accurate shift to the rest frames with the precision of \textbf{the} LASP radial velocities. 

\subsection{Flux Calibration and Dereddening}

Before co--adding of the spectra which were observed at different times,  it is necessary to correct the dereddening errors associated with the LAMOST flux calibration. For the LAMOST flux calibration, the LAMOST 2D pipeline picks out several high--SNR standards in the temperature range  of 5750--7250 K and then obtains the spectrograph response curve (SRC) by comparing the observed spectra with synthetic spectra (using the corresponding parameters from the KURUCZ spectral library). As a result, the dereddening uncertainties of standards, particularly for stars in high dense fields,  have an impact on the SRC derivation.  This introduces uncertainties to all spectra of the spectrograph for one plate. For a few spectra ($\sim$ 2\%), flux calibrations are completed by using the average response curve of each spectrograph (ASPSRC) method \citep{du16}. This method has some uncertainties casued by variations in the shape of individual SRCs, which introduces uncertainties to the  calibrated spectra. Because the stellar parameters of the observed spectrum have been well determined, we recalibrated each spectrum by comparing the observed spectra with  synthetic spectrum,  using the corresponding parameters from the KURUCZ spectral library \citep{castelli}.

\subsection{Co--adding}

We selected and co--added spectra in bins of $T_{\rm eff}$, log \emph{g},  and [Fe/H] to create our empirical catalog. Considering both errors,  external comparisons and the internal validation as the ELODIE interpolator was applied to the spectra at the LAMOST resolution, we separated spectra into the following  parameter bins:  $T_{\rm eff}$ in steps of 150 K,  log \emph{g} in steps of 0.25 dex,  and [Fe/H] in steps of  0.15 dex.   We applied some basic quality cuts in order to ensure the quality of the spectra involved in the co--adding.  We removed  the low--SNR (SNRg \textless 10.0) and large RV error spectra, in which the RV error $>$ 15.0 ${\rm km \, s}^{-1}$,  when we selected data from the sql database. We calculated the instrumental full--width--half--maximum ( FWHM ) of the blue and red arms from the arc lamps of LAMOST DR5. We removed the spectrum of arc lamp having FWHMb or FWHMr  outside of  3$ \sigma $ of their distributions.  The LAMOST spectra are logarithmically spaced and in vacuum wavelengths. Using the radial velocities determined  by LASP,  we shifted all of the spectra into their rest frames. We re--sampled all of the spectra to a set of fixed wavelengths for alignment over their wavelengths. The wavelength--justified spectra were then normalized to a median value of unity, where the spectrum was divided by the median of its flux values,  and  were co--added  by using a statistical method to obtain a reliable flux at each wavelength element (equivalent to clustering). For groups with more than 5000 spectra, only the first 5000 spectra with the largest SNR were selected.  Finally, we trimmed the flux grids at 3800 \AA \ and 8900 \AA \ to avoid areas that were not complete after the radial velocity shifts. In addition, we calculated the standard deviation of the co--added templates at each wavelength element for each template.

\subsection{Quality control}
We initially obtained a grid of  3301 templates by co--adding.  Each grid node included the ranges of stellar parameters ($T_{\rm eff}$, log \emph{g},  and [Fe/H])  and the corresponding co--added spectrum. We used LASP to estimate the stellar parameters from the co--added spectra. For K giants, we calibrated the LASP--derived values of log \emph{g} using the relations given by Eq \ref{eq:eq1} and Eq \ref{eq:eq2}. The agreement of parameters between the co--added spectrum and the individuals used for co--adding validated the template grid node. The co-added spectra totaled 2996 with estimated parameters falling into the previous bins of $T_{\rm eff}$, log \emph{g}, and [Fe/H]. We manually inspected the 2996 spectra and removed 178  unreliable spectra.  In addition, we manually inspected the other 305 co--added spectra and  incorporated into the final template grid 74 spectra having parameters that deviated little from those in previous bins.  We ultimately obtained  a grid of 2892 templates.

\section{Results} \label{sec:result}
\begin{figure}
\begin{minipage}{0.5\linewidth}
  \centerline{\includegraphics[width=8cm]{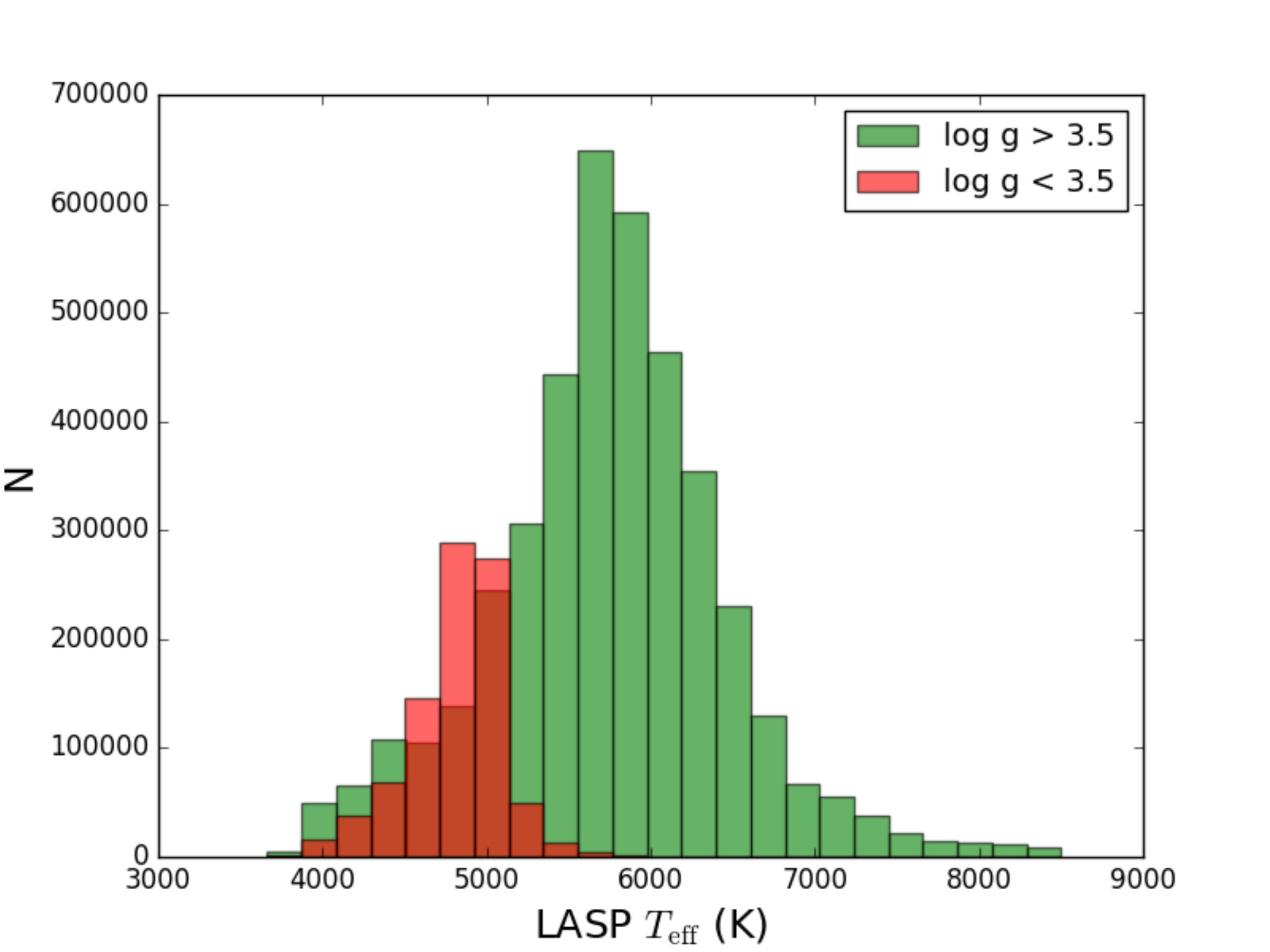}}
\end{minipage}
\begin{minipage}{0.5\linewidth}
  \centerline{\includegraphics[width=8cm]{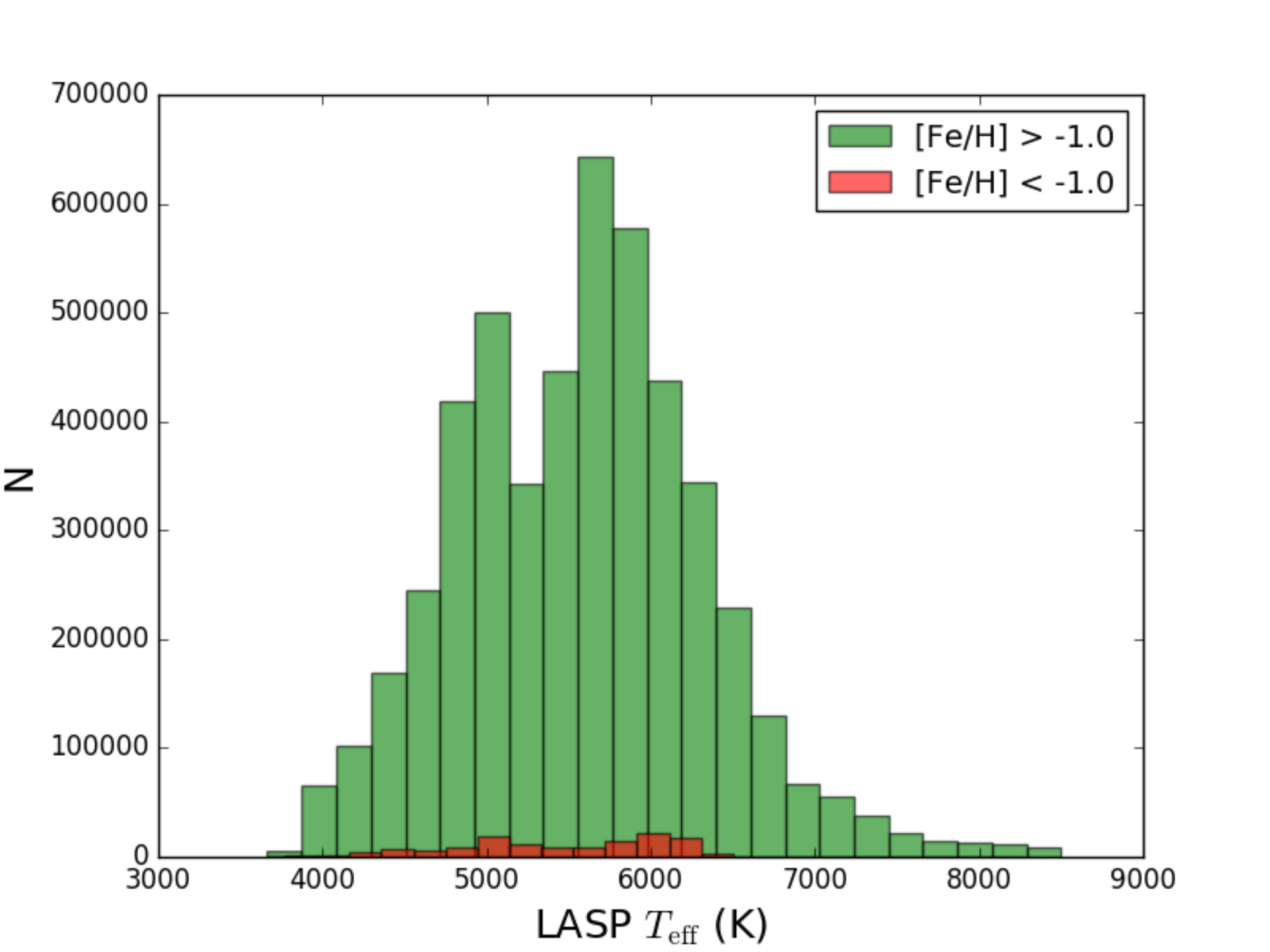}}
\end{minipage}
\caption{Histograms of the numbers of stars in the temperature space. On the left, histograms for roughly dwarfs (giants) are shown in green (red). On the right, plots for [Fe/H] $>$ -1.0 dex ($<$-1.0 dex) are shown in green (red).}
\label{fig2}
\end{figure}

More than 5 million spectra with well--determined stellar parameters  were included in  LAMOST DR5. We obtained $\sim$ 4.6 million spectra to create this library  after applying some basic quality cuts. The distributions of the numbers of stars in the temperature space are shown in  Fig \ref{fig2}. We determined that LAMOST has obtained  almost the same number of K type stars as those of F and G type stars.  We noticed an abundance of K giant stars  and a scarcity of metal--poor stars.

\subsection{The stellar parameter coverage}

\begin{figure}
\begin{minipage}{0.5\linewidth}
  \centerline{\includegraphics[width=8cm]{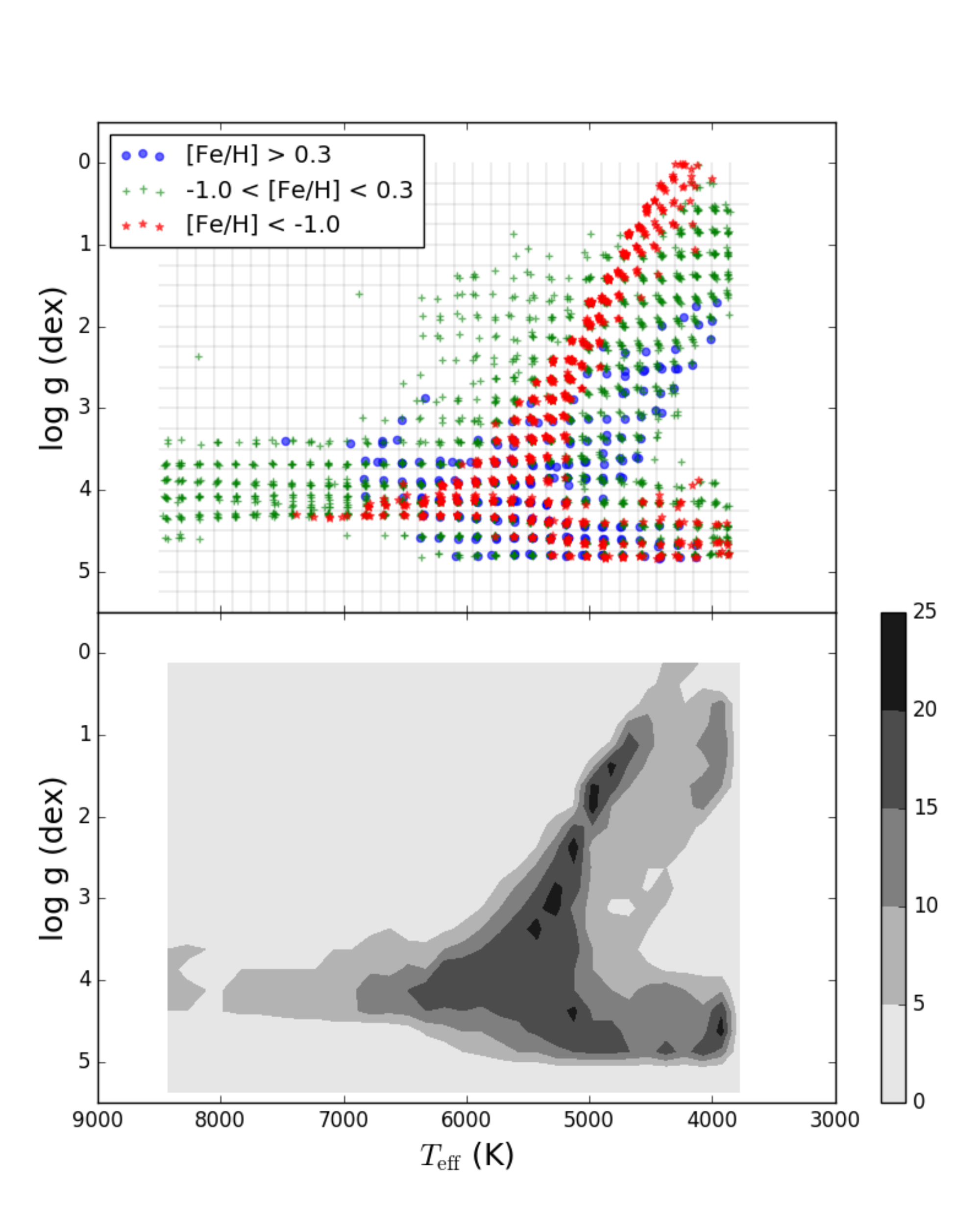}}
\end{minipage}
\begin{minipage}{0.5\linewidth}
  \centerline{\includegraphics[width=8cm]{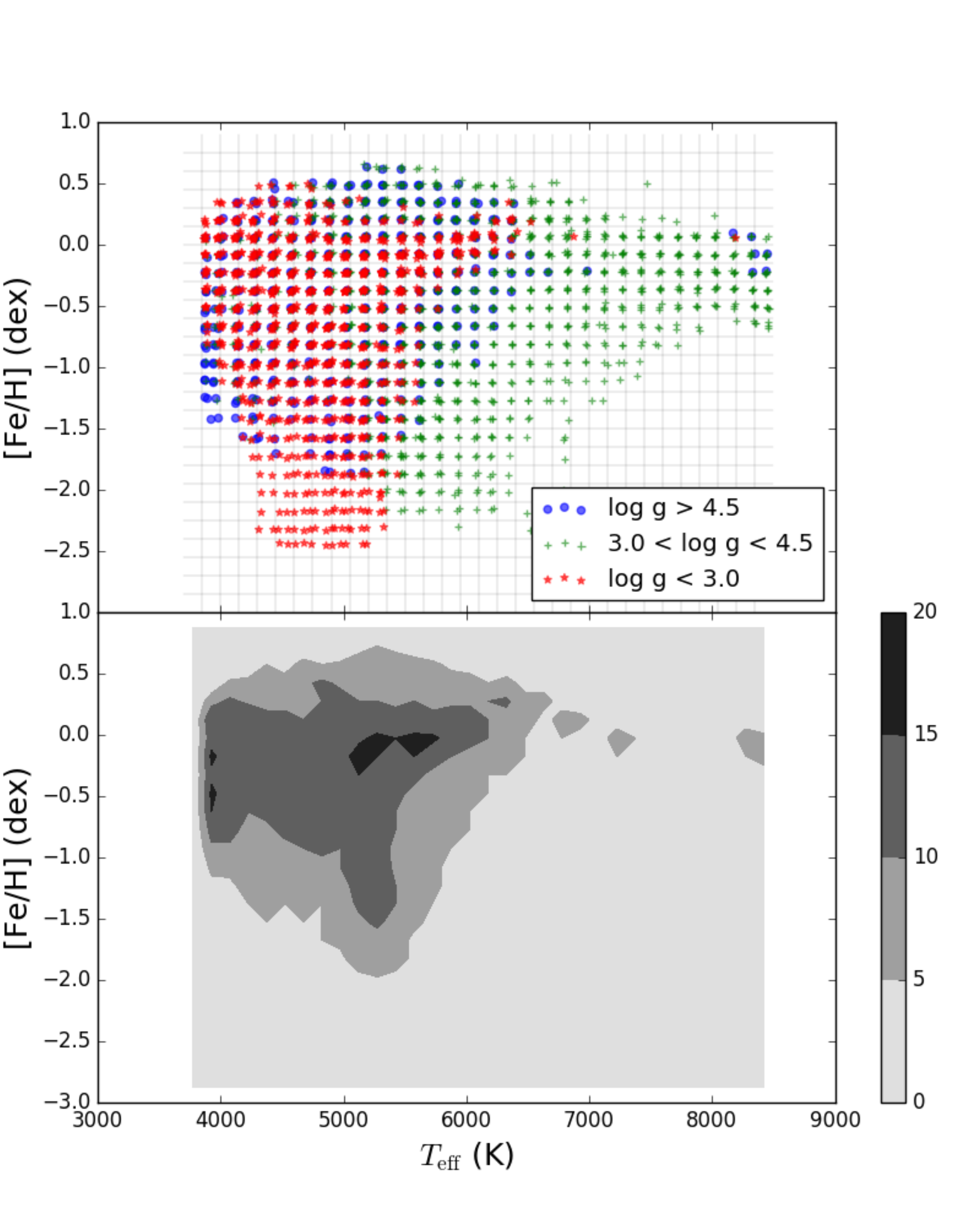}}
\end{minipage}
\caption{Distributions of our library in the $T_{\rm eff}$--log \emph{g}  and $T_{\rm eff}$--[Fe/H] planes of the labelled parameters. In the top left right panel, the symbol color distinguishes different  metallicity (surface gravity) classes.  The bottom panel shows a contour map of the numbers of stars in the  $T_{\rm eff}$--log \emph{g} (left) and $T_{\rm eff}$--[Fe/H] (right) planes with grid steps of (150 K, 0.25 dex) and (150 K, 0.15 dex), respectively. }
\label{fig3}
\end{figure}

The distribution of stellar parameters in each grid  was approximately  uniform, and the mean and standard deviation corresponded to parameters and parameter errors of the grid template, respectively. We statistically determined the stellar parameters and the corresponding errors for the 2892 template spectra. By construction, our library contains high--quality template spectra that span a large region of the H--R diagram, at $T_{\rm eff} \approx  $ 3700 -- 8500 K, log \emph{g} $\approx$ 0 -- 5.0 dex  and [Fe/H] $\approx$ -2.5 -- 1.0 dex , with more additions of cool star spectra.  Fig \ref{fig3} shows the domain of $T_{\rm eff} $, log \emph{g},  and [Fe/H].  We noticed that our library includes more K type stars,  and the stars are evenly distributed in the parameter space. Therefore, this empirical library offers the most complete empirical spectra of K type stars.

\clearpage

\subsection{The Spectra}

\begin{figure}
\centering
\includegraphics[width=16.0cm,height=18.0cm]{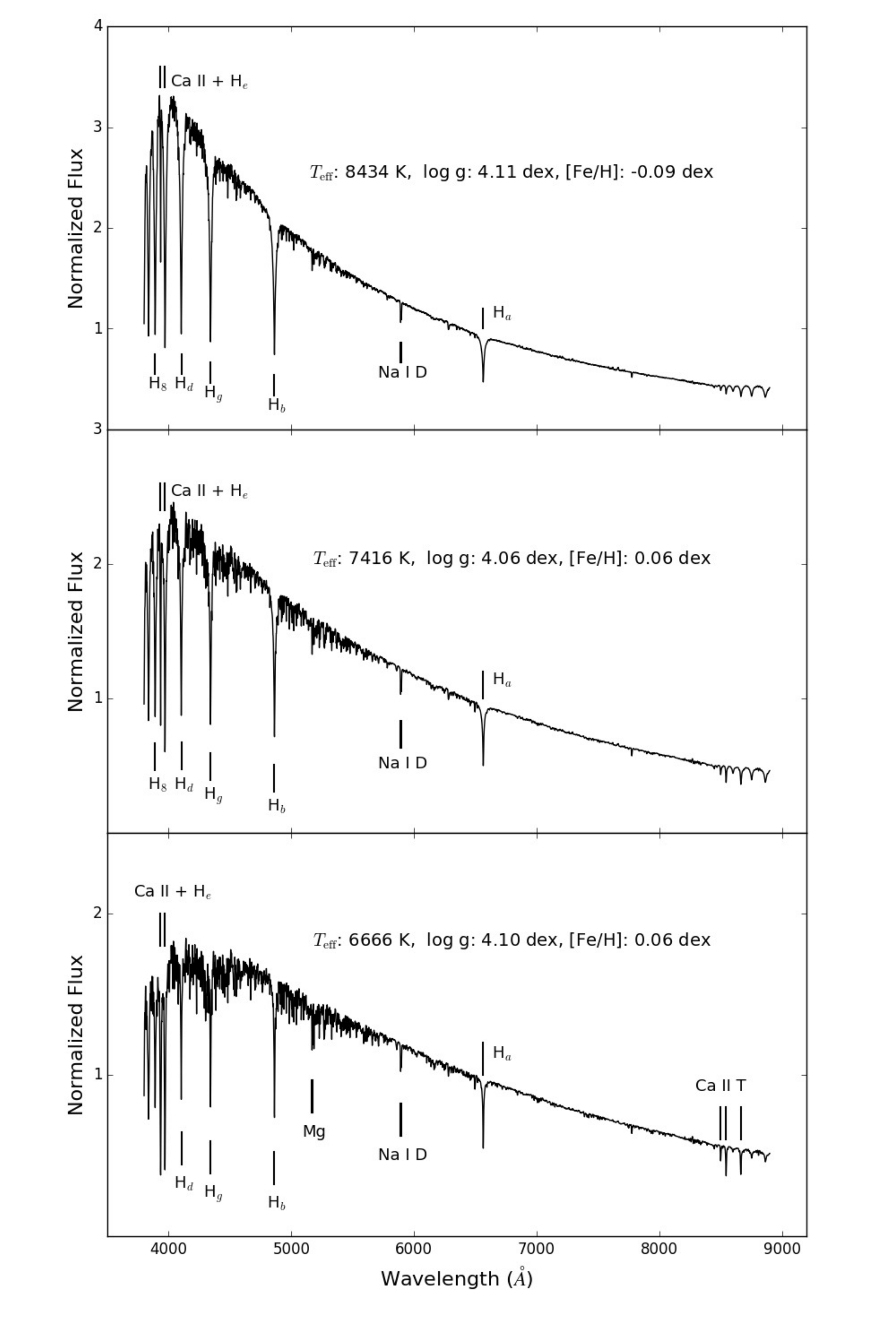}
\caption{Examples of template spectra from  main--sequence, high--temperature with roughly solar metallicity.  All of the prominent absorption features are labeled for each spectrum:  H$_{a}$  for H$\alpha$,  H$_{b}$  for H$\beta$,  H$_{g}$ for H$\gamma$,  H$_{d}$ for H$\delta$, and H$_{e}$ for H$\epsilon$.}
\label{fig4}
\end{figure}

\begin{figure}
\centering
\includegraphics[width=16.0cm,height=18.0cm]{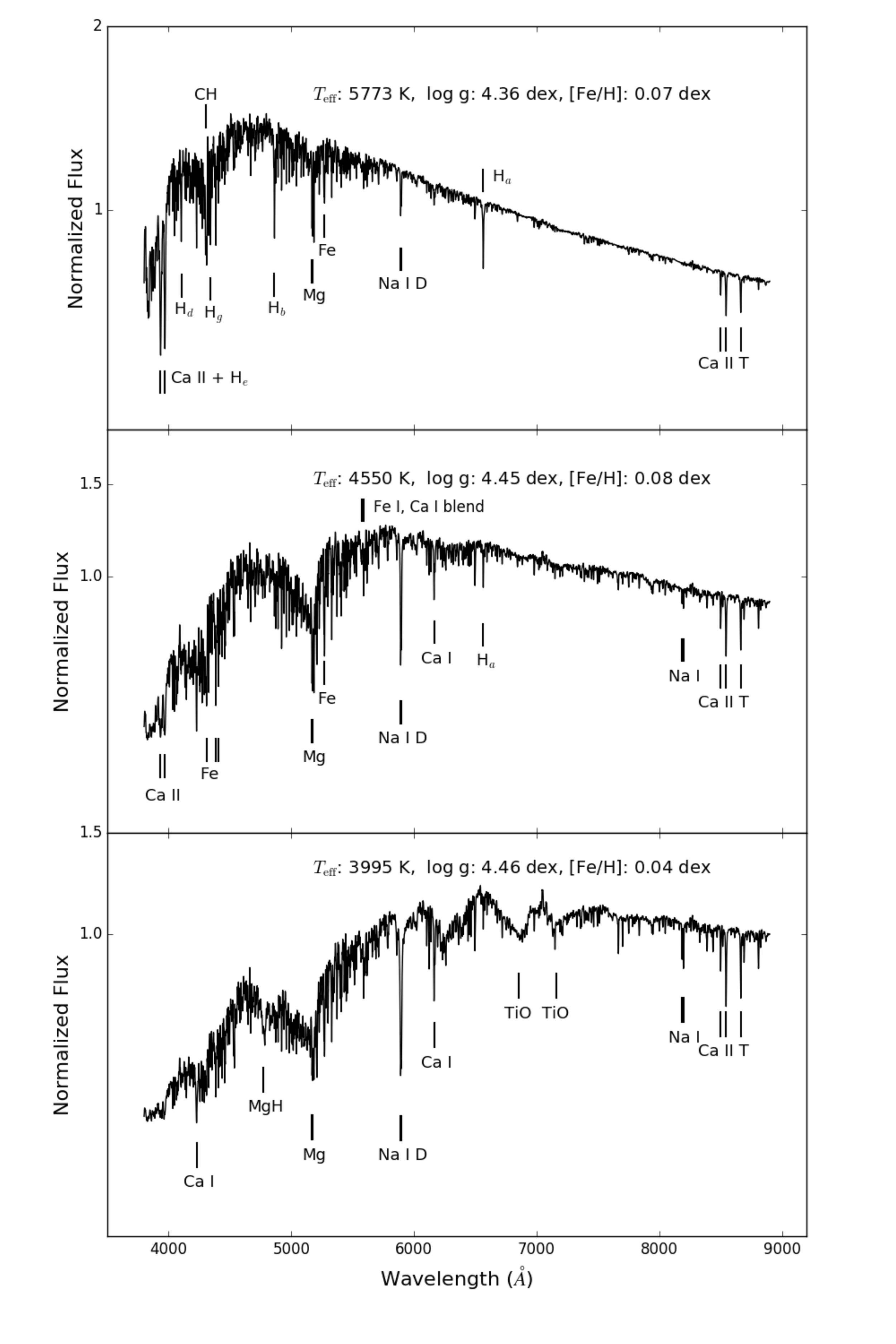}
\caption{ Examples of template spectra from main-sequence, low--temperature with roughly solar metallicity.  All of the prominent absorption features are labeled for each spectrum:  H$_{a}$  for H$\alpha$, H$_{b}$ for H$\beta$,  H$_{g}$ for H$\gamma$,  H$_{d}$ for H$\delta$,  and H$_{e}$ for H$\epsilon$.}
\label{fig5}
\end{figure}

\begin{figure}
\centering
\includegraphics[width=16.0cm,height=18.0cm]{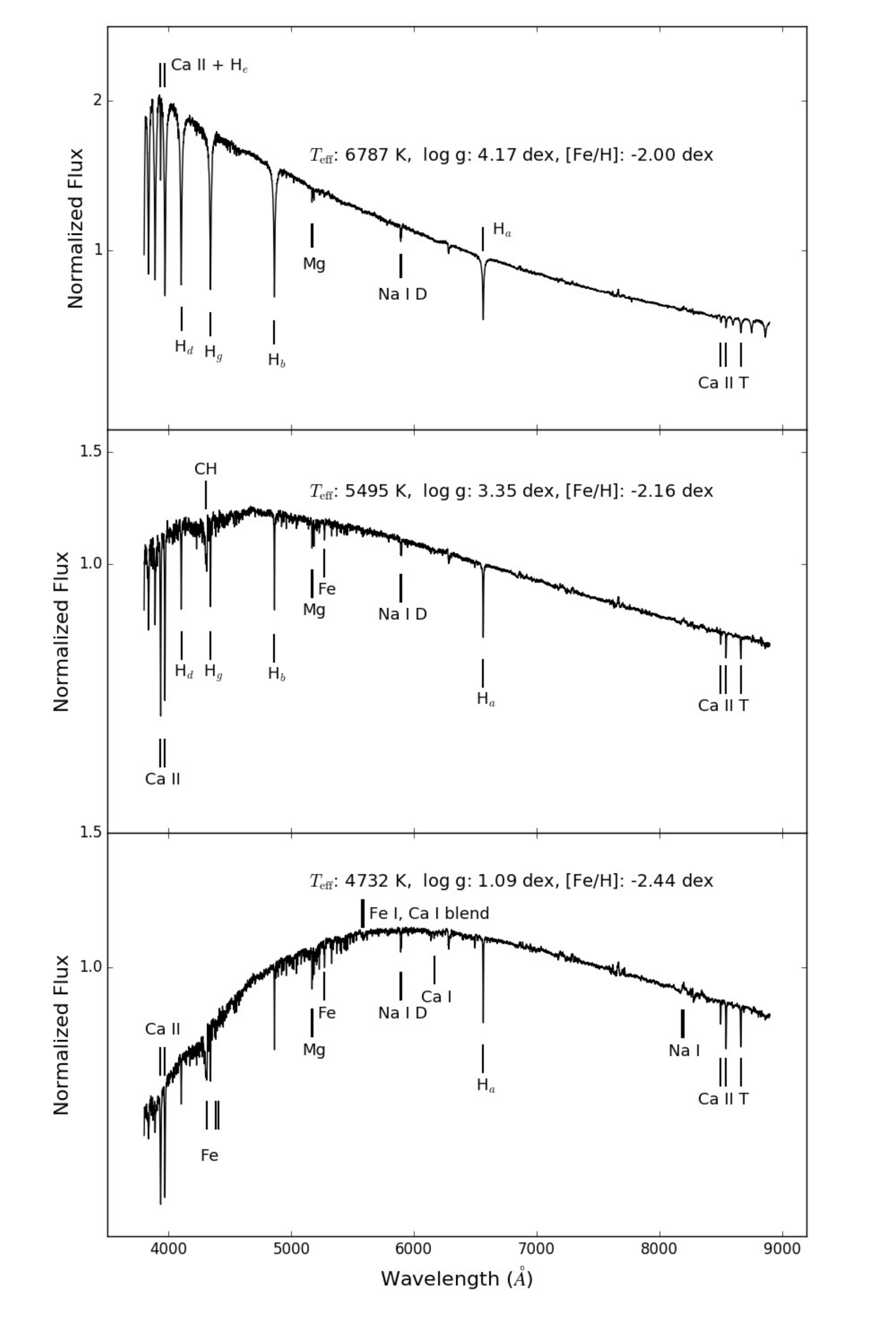}
\caption{ Example spectra of low metallicity stars from three  $T_{\rm eff}$--log \emph{g} grid nodes.  All of the prominent absorption features are labeled as in Fig \ref{fig5} for comparison: H$_{a}$  for H$\alpha$, H$_{b}$ for H$\beta$,  H$_{g}$ for H$\gamma$,  H$_{d}$  for H$\delta$,  and H$_{e}$ for H$\epsilon$.}
\label{fig6}
\end{figure}

\begin{figure}
\begin{minipage}{0.5\linewidth}
  \centerline{\includegraphics[width=8cm]{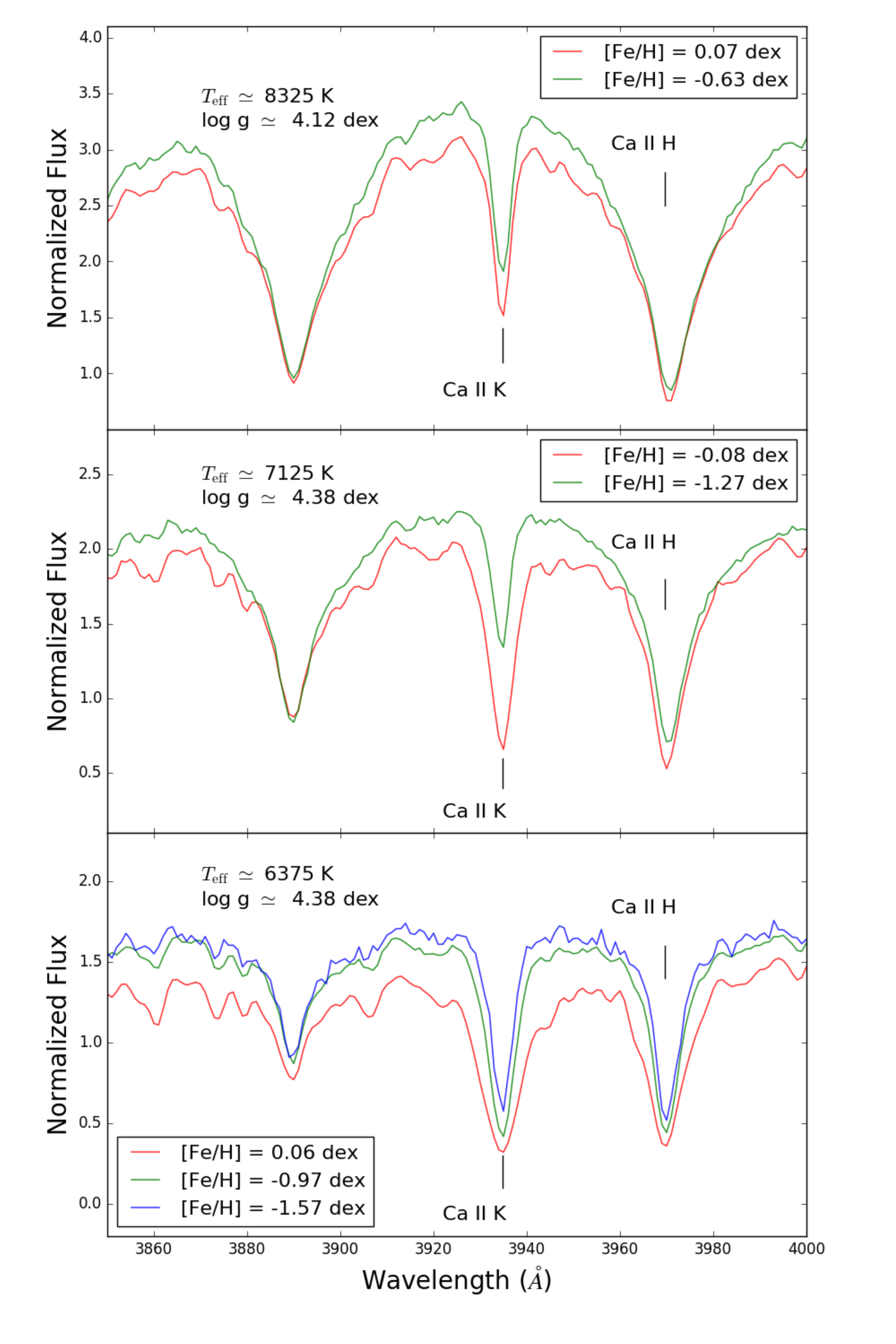}}
\end{minipage}
\begin{minipage}{0.5\linewidth}
  \centerline{\includegraphics[width=8cm]{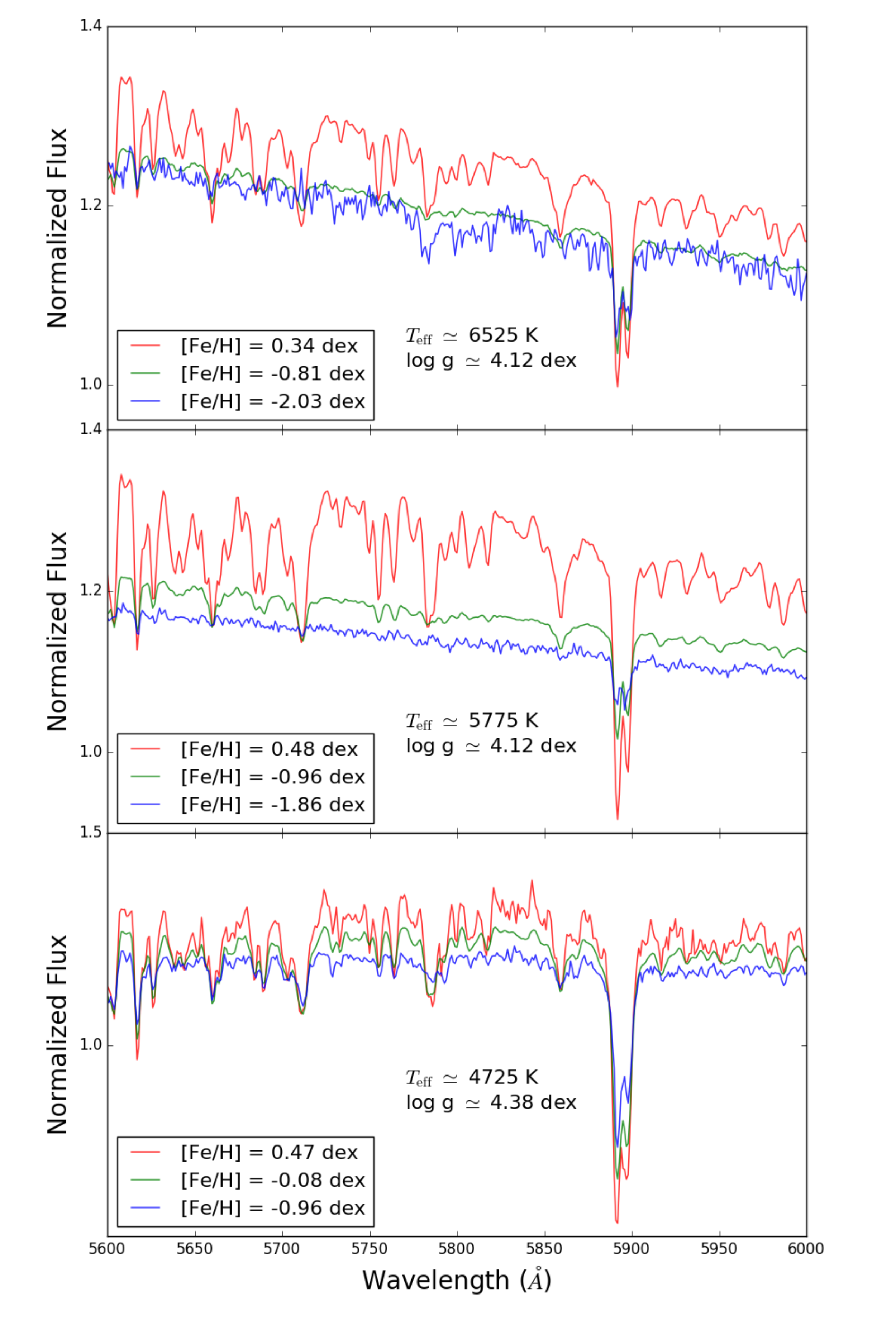}}
\end{minipage}
\caption{Metallicity variations for a range of temperatures, shown in two different wavelength regions. The left panel shows the region around the Ca II H and Ca II K lines ($\sim$ 3940 \AA ), which are metallicity sensitive for the higher--temperature stars ( $\sim$ 6200--8500 K). The right panel shows the region around the Na I D lines ($\sim$ 5900 \AA), which is metallicity--sensitive for the spectral subclasses of around F through late K--type stars ($\sim$ 3600--7000 K).}
\label{fig7}
\end{figure}

We obtained a set of 2892 template spectra. All of the spectra are available online in the FITS formats\ \footnote{\url{http://paperdata.china-vo.org/empirical-lib/LAMOST-Empirical-lib/LAMOST-Emp-library.tar.gz}}.  Fig \ref{fig4} and Fig \ref{fig5}  show one spectrum from each temperature grid node near solar metallicity, with the prominent absorption features labeled.  Fig \ref{fig6} shows low metallicity spectra from three $T_{\rm eff}$--log \emph{g} grid nodes, with the absorption features labeled as in Fig \ref{fig5} for comparison.  The roughly solar--metallicity spectra clearly have  deeper absorption features than the low metallicity templates. In the latter, some absorption lines that are sensitive to metallicities weakened or even disappeared. For the low metallicity templates ([Fe/H] $\le$ -1.5 dex ),  very few low metallicity spectra were co-added to create the templates (see Fig \ref{fig2}).  Therefore, the low metallicity spectra are generally noisier than their higher metallicity counterparts, although they still show a lack of real absorption features in our templates. Fig \ref{fig7} shows an enlarged view of two absorption features that are sensitive to metallicity. The left panel shows the region around the Ca II H and Ca II K lines ( $\sim$ 3940 \AA ) and the right panel shows that around the Na I D lines ($\sim$ 5890 \AA ).  The Ca II H and Ca II K lines are metallicity sensitive for high--temperature  A and F stars, whereas the Na I D lines are metallicity sensitive for low--temperature, around F through early M--type stars \citep{kesseli}.  The Ca II K line can be used as a metallicity indicator for A and F stars and the Na I D lines are a useful metallicity indicator and also a gravity indicator for F through late K--type stars. The equivalent widths of both of these features are known to be larger in the  higher--metallicity spectra, implying deeper absorption features, than in their lower--metallicity counterparts.  This result is confirmed in our templates.

\begin{figure}
\centering
\includegraphics[width=16.0cm,height=18.0cm]{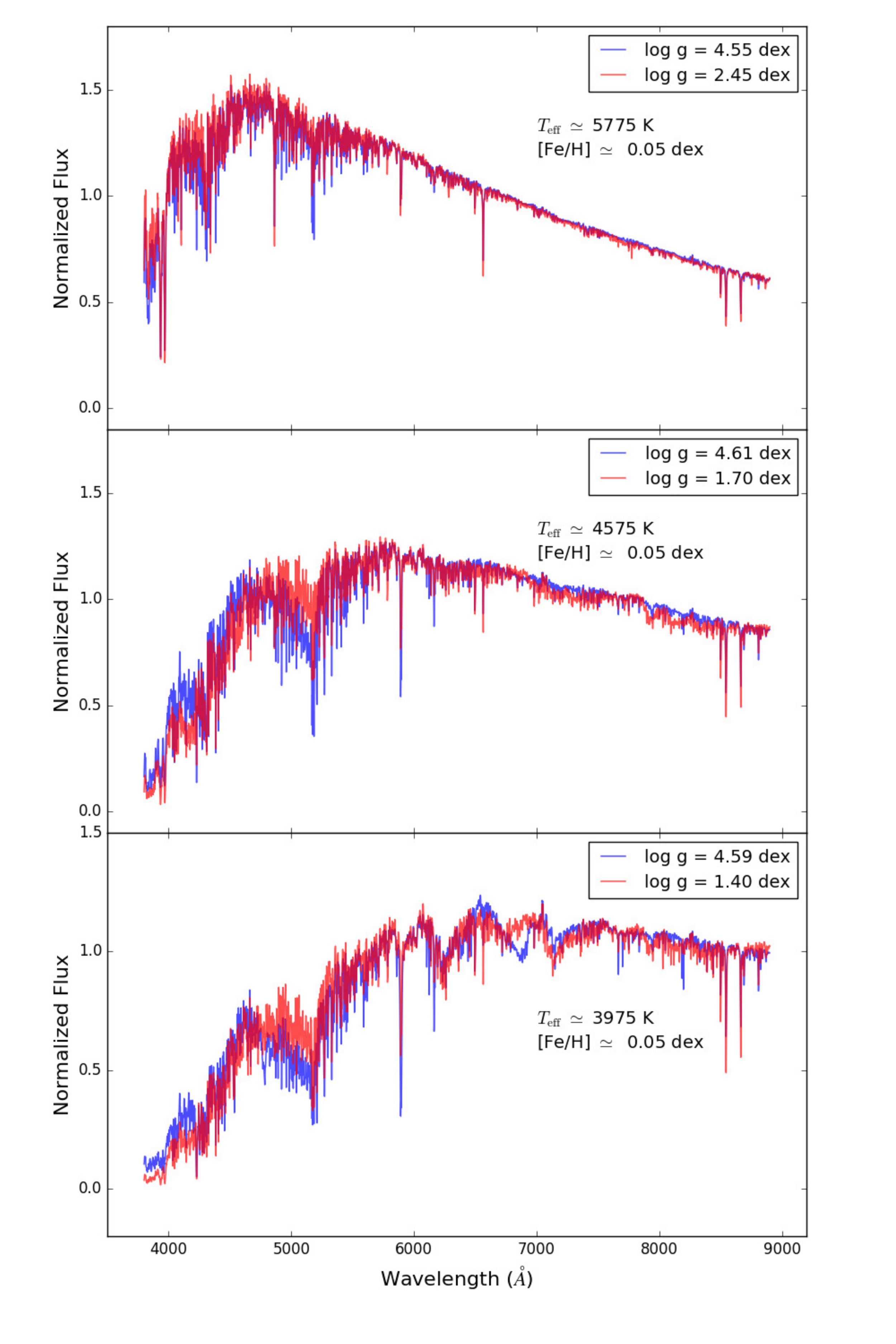}
\caption{ Surface gravity comparison between dwarf and giant templates of the same metallicity and temperature for the entire spectrum. The blue (red) line shows the dwarf (giant) template.}
\label{fig8}
\end{figure}

\begin{figure}
\centering
\includegraphics[width=16.0cm,height=16.0cm]{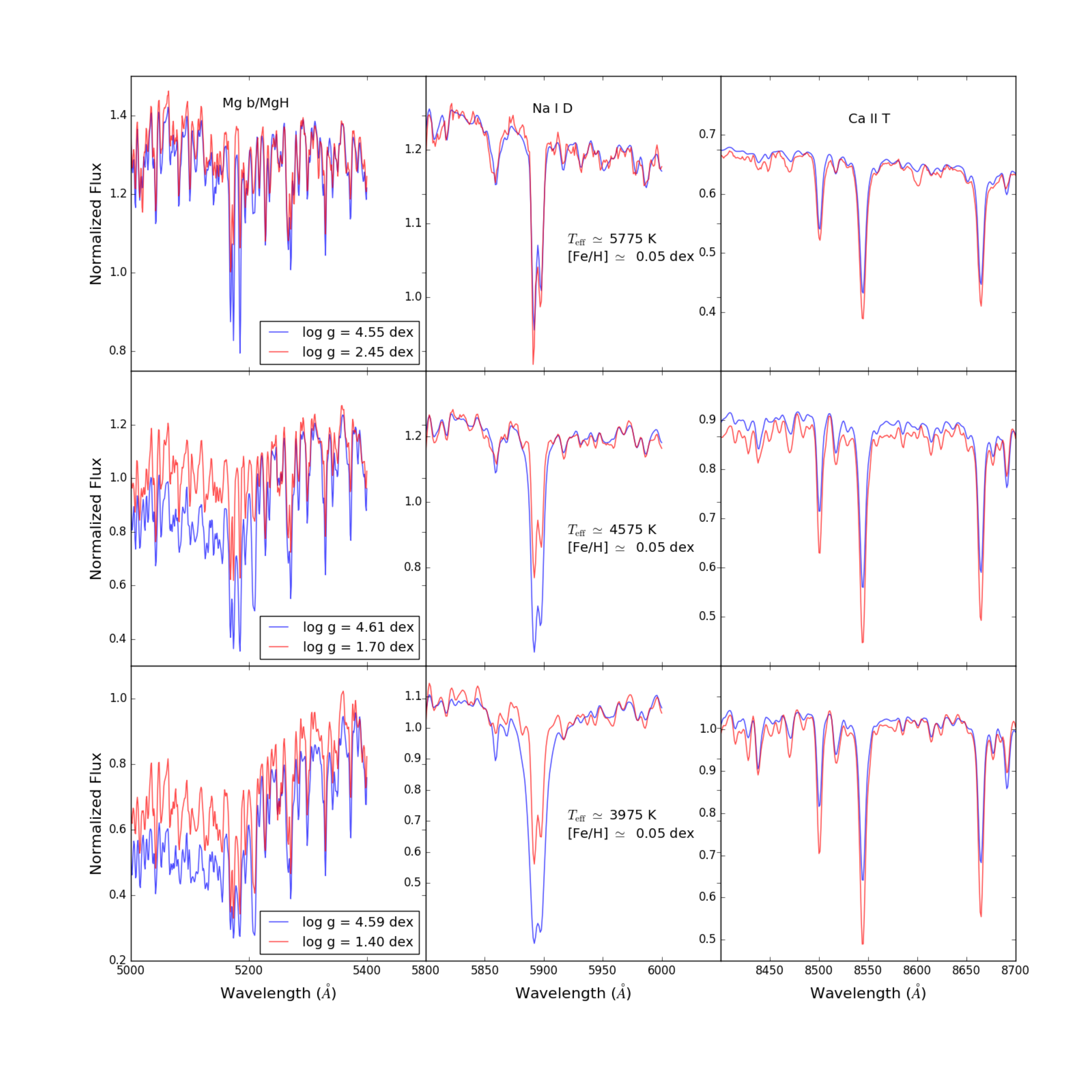}
\caption{Surface gravity comparison between dwarf and giant templates of  Fig \ref{fig8} shown in the expanded regions. The left panel shows the region around the Mg b/MgH feature ($\sim$ 5200 \AA ), the middle panel shows the region around the Na I D lines ($\sim$ 5900 \AA), and the right panel shows an expanded region around the Ca II triplet. All of these lines are known to be sensitive to log \emph{g}, which was confirmed in our templates.}
\label{fig9}
\end{figure}

Fig \ref{fig8} and Fig \ref{fig9} show differences between dwarf and giant spectra for a range of temperatures. Both the dwarf and the giant stars of each pair show the same metallicity and temperature.  Fig \ref{fig8} shows examples of the entire spectrum for both a dwarf and a giant spectrum for the temperatures of 5775 K ( $\sim$  G5), 4575 K ( $\sim $ K5 ),  and 3975 K ( $\sim $ K7).  Fig \ref{fig9} shows an expanded view of  gravity sensitive features in different regions of the template spectra ( left panel: Mg b/MgH; middle panel:  Na I D;  right panel: Ca II triplet ). The Mg b and MgH feature ( $\sim $ 5200 \AA) is prominent in dwarfs, weaker in Population I giants as compared to dwarfs, and is absent in metal--poor giants \citep{helmi}. The Na I D lines are extremely strong in dwarfs and weak in giants, therefore, this feature is  often used as an gravity indicator  \citep{kesseli}.  However,  the Na I D lines located  at the junction of the blue and red arms, are sometimes not credible owing to the low sensitivity of the LAMOST instrument.  Finally,  the giants have deeper absorption features and thus larger equivalent widths for the Ca II triplet ( $\sim$ 8600 \AA ) as compared to dwarfs \citep{kesseli}, as shown in the right panel of Fig \ref{fig9}.

\clearpage

\clearpage

\clearpage

\clearpage

\clearpage

\clearpage

\begin{figure}
\begin{minipage}{0.5\linewidth}
  \centerline{\includegraphics[width=8cm]{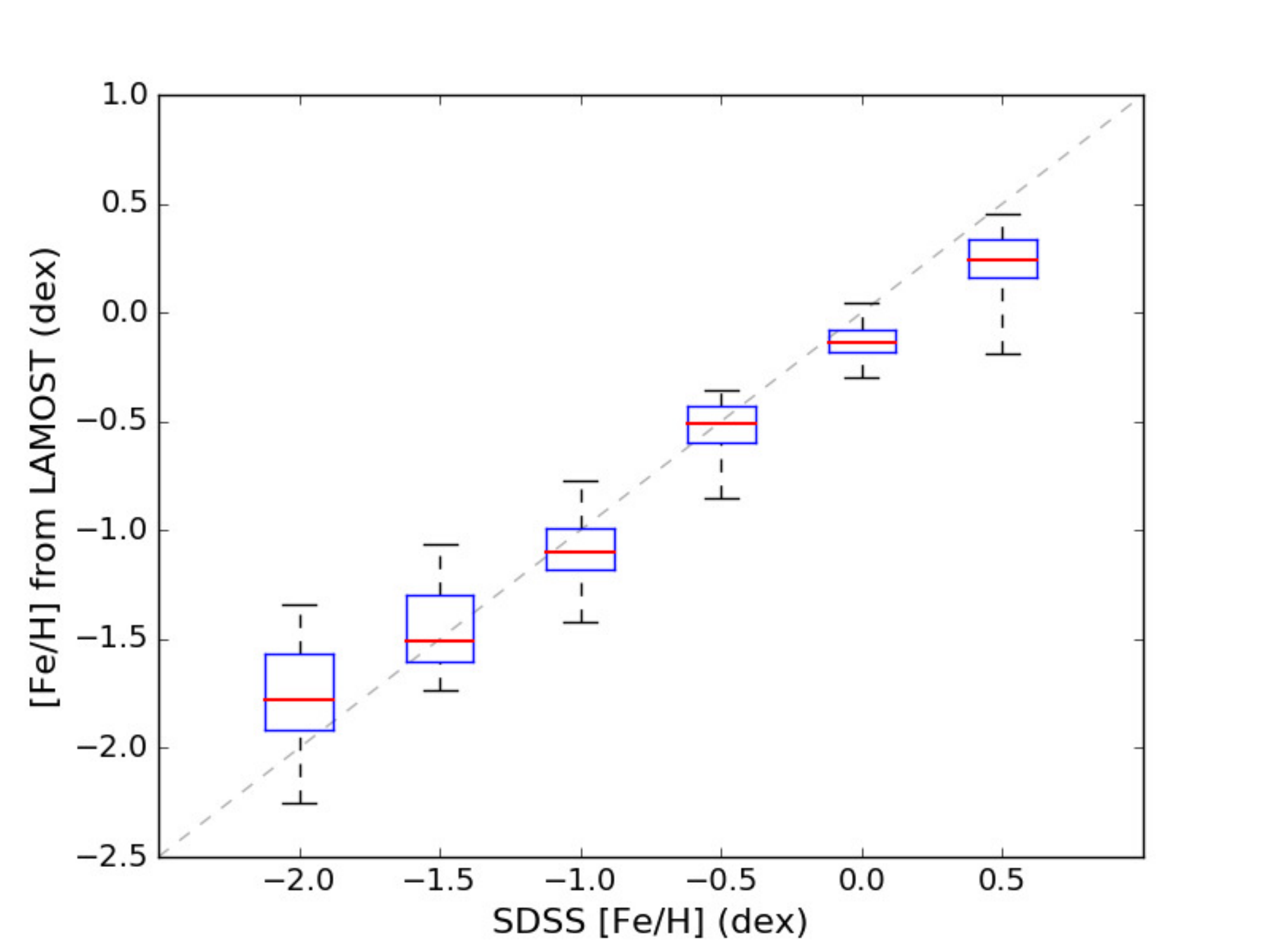}}
\end{minipage}
\begin{minipage}{0.5\linewidth}
  \centerline{\includegraphics[width=8cm]{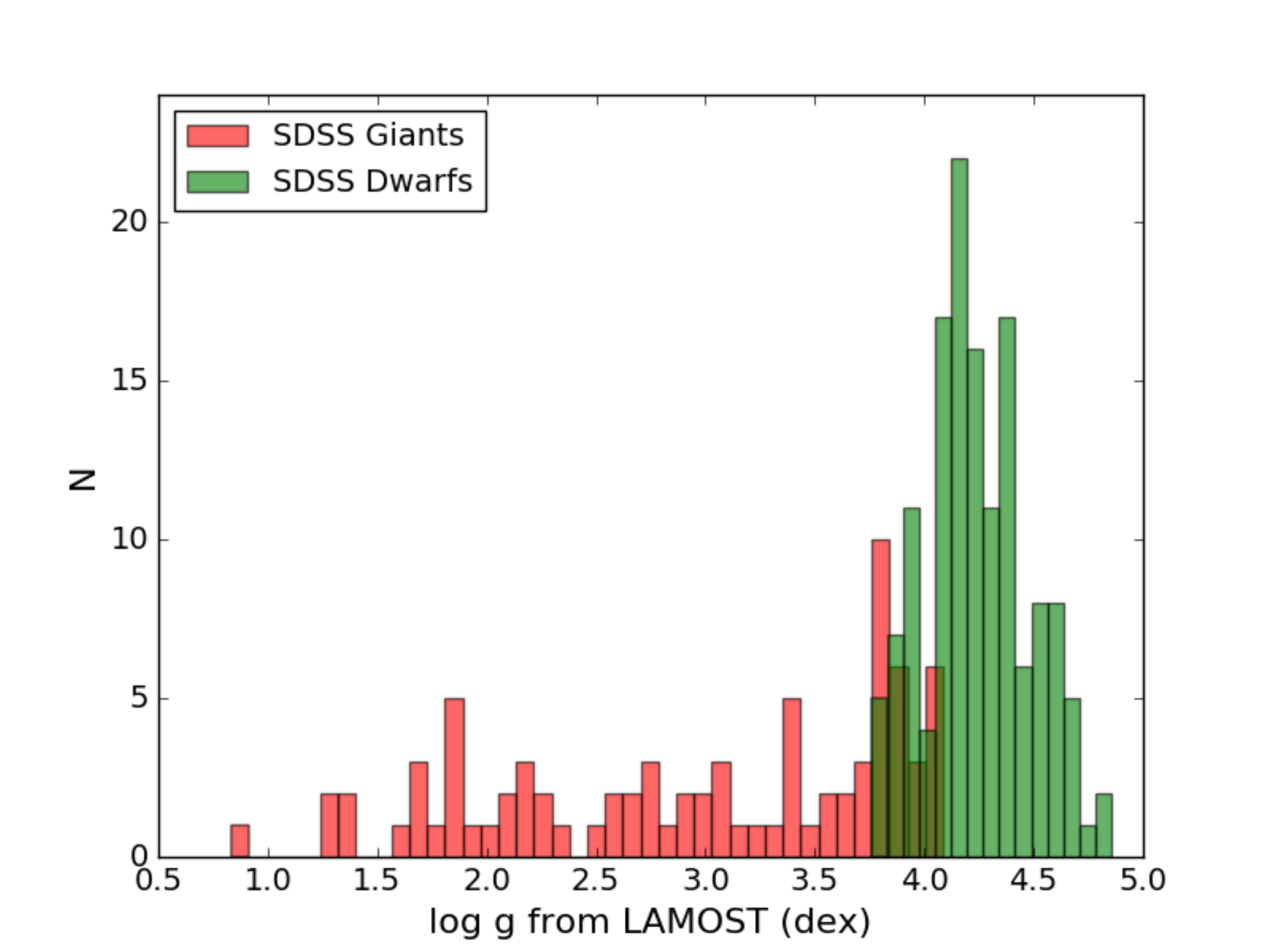}}
\end{minipage}
\caption{Comparisons of SDSS and LAMOST libraries.  The left side shows comparisons of the SDSS metallicities and those estimated by comparing the SDSS spectra to LAMOST library. The values of the horizontal axis are the SDSS library metallicities,  and the values of the vertical axis are the distributions of their metallicities estimated from the LAMOST library. The box extends from the lower to upper quartile values of the derived metallicities, with a line at the median. The whiskers extend from the box to show the metallicity range. On the right,  histograms of surface gravities determined  by comparing the SDSS spectra to the LAMOST template spectra. Histograms for dwarfs (giants) are shown in green (red) color.} 
\label{fig10}
\end{figure}

\begin{figure}
\centering
\includegraphics[width=16.0cm,height=18.0cm]{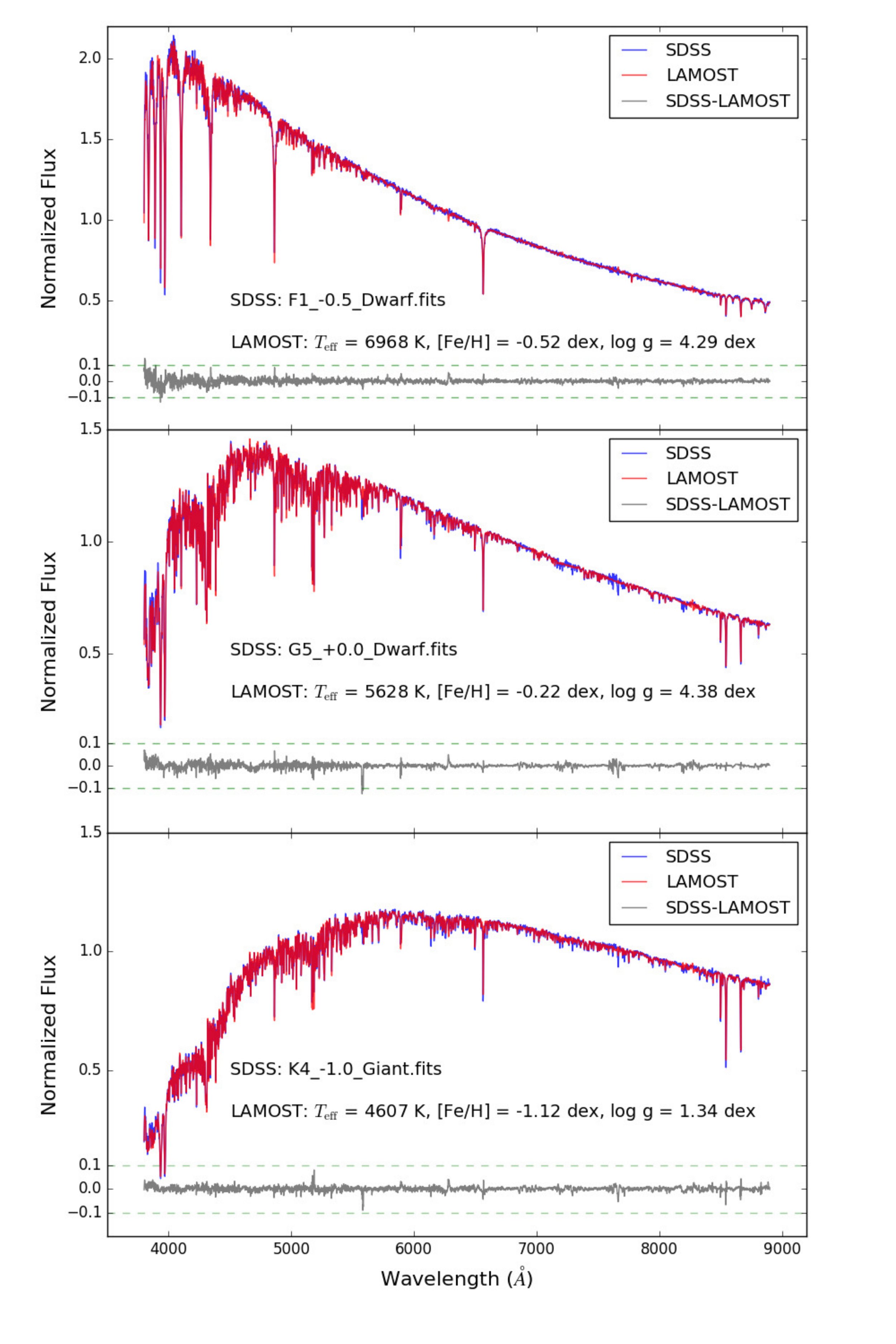}
\caption{Comparisons of the SDSS template spectra  and the best--matched  LAMOST template spectra.  The blue (red) line shows the SDSS template (the LAMOST template), and the gray line shows the flux differences between them.  The +0.1 and -0.1 of the vertical axis are also marked by the green dashed lines.}
\label{fig11}
\end{figure}

We compared our template spectra with the empirical library created using spectra from  SDSS \citep{kesseli}. The SDSS library includes spectral classes (O5 through L3), luminosity classes (Dwarf or Giant),  and metallicities(-2.5 dex, -2.0 dex, -1.5 dex, -1.0 dex, -0.5 dex, +0.0 dex, +0.5 dex).  To complete a thorough comparison, we estimated the stellar parameters from the SDSS template spectra by comparing  SDSS spectra to our library, using  a $ \chi^2 $   minimization in multidimensional parameter space.  We did not estimate the stellar parameters of SDSS spectra that were not in our stellar parameter space.  Fig \ref{fig10} (left) shows the distributions of  metallicities estimated by comparing the SDSS spectra to LAMOST library. The LAMOST--derived metallicities agreed well  with the SDSS metallicities for the abundance patterns available within the solar neighborhood. This occurred because both LAMOST and SDSS have obtained sufficient samples of stars close to the Sun. However, the lack of adequate low metallicity stars presents a limitation.  Thus, the differences between the SDSS  metallicities and the LAMOST--derived metallicities are greater for low metallicity stars when compared with others. The metallicities of metal--poor stars,  such as [Fe/H] $=$ -2.0 dex,  are at the boundary of the metallicity coverage of our library.  Target stars must necessarily match stars in the interior of  the parameter space of the reference library,  as a result, the LAMOST--derived metallicities for the SDSS stars of [Fe/H] $=$ -2.0 dex are pulled toward richer metallicities. The same problem  exists in the determinations of surface gravities of SDSS hot giants. We are limited by the numbers of hot giant stars, including F and early--type  G  giants (see Fig \ref{fig3}),  the matches in the interior of the parameter space result in the LAMOST--derived surface gravities of SDSS hot giants  being pulled toward dwarfs, as shown in Fig \ref{fig10} (right). For SDSS cool giants, including late-type G and K, the LAMOST--derived surface gravities confirm that they are giants. For dwarfs classified by the SDSS pipeline,  the LAMOST--derived surface gravities agree with the SDSS luminosity classes.  Fig \ref{fig11} shows the comparisons of the SDSS template spectra and the best--matched  LAMOST template spectra. We adopted a multiplicative polynomial to absorb the differences in their flux calibrations. The LAMOST template spectra agree well with the  SDSS template spectra,  and their flux residuals are negligible (See  Fig \ref{fig11} ).

\clearpage

\subsection{ Individual Spectral Feature Lines }

\begin{figure}
\begin{minipage}{0.5\linewidth}
  \centerline{\includegraphics[width=8cm]{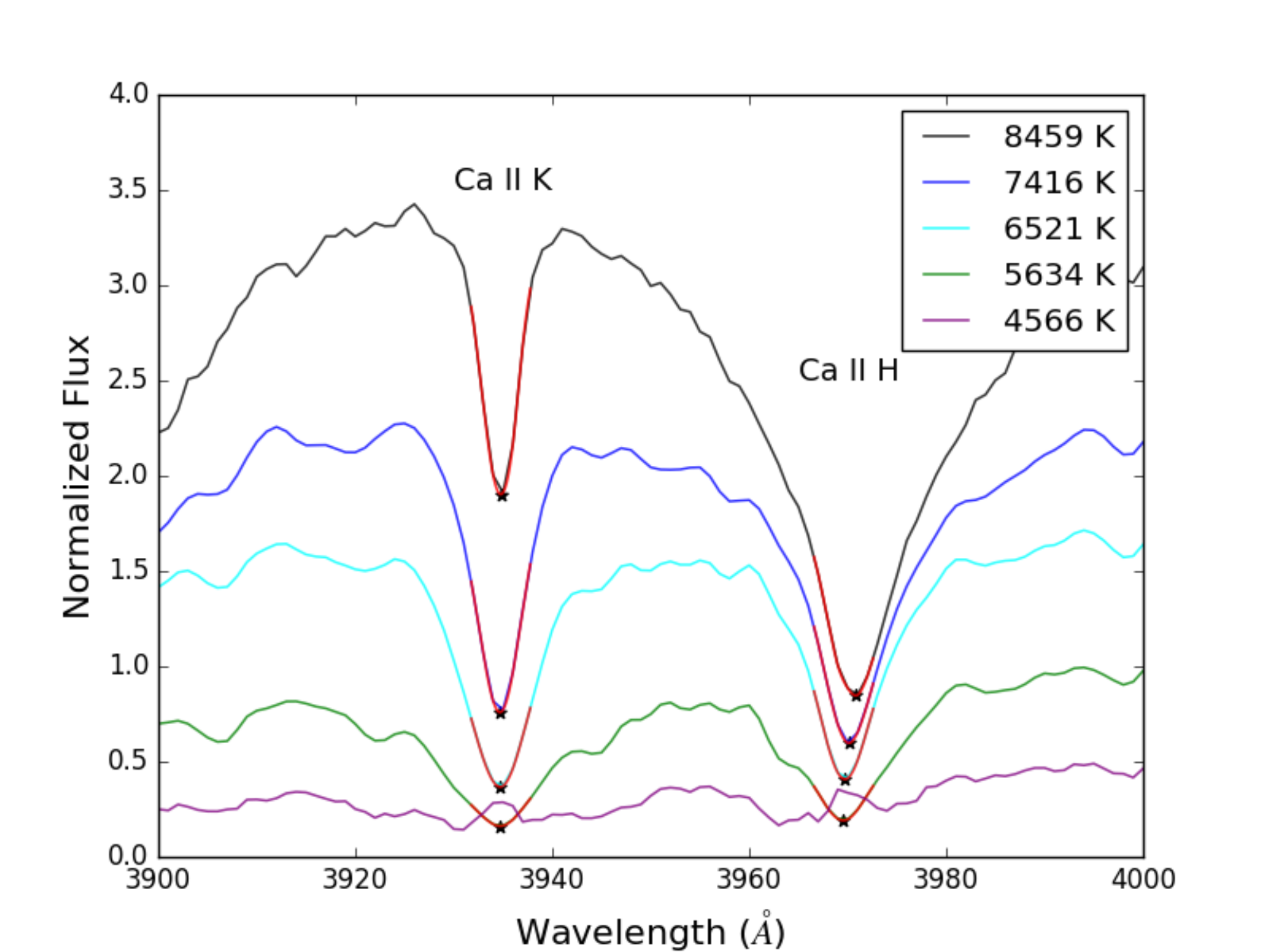}}
\end{minipage}
\begin{minipage}{0.5\linewidth}
  \centerline{\includegraphics[width=8cm]{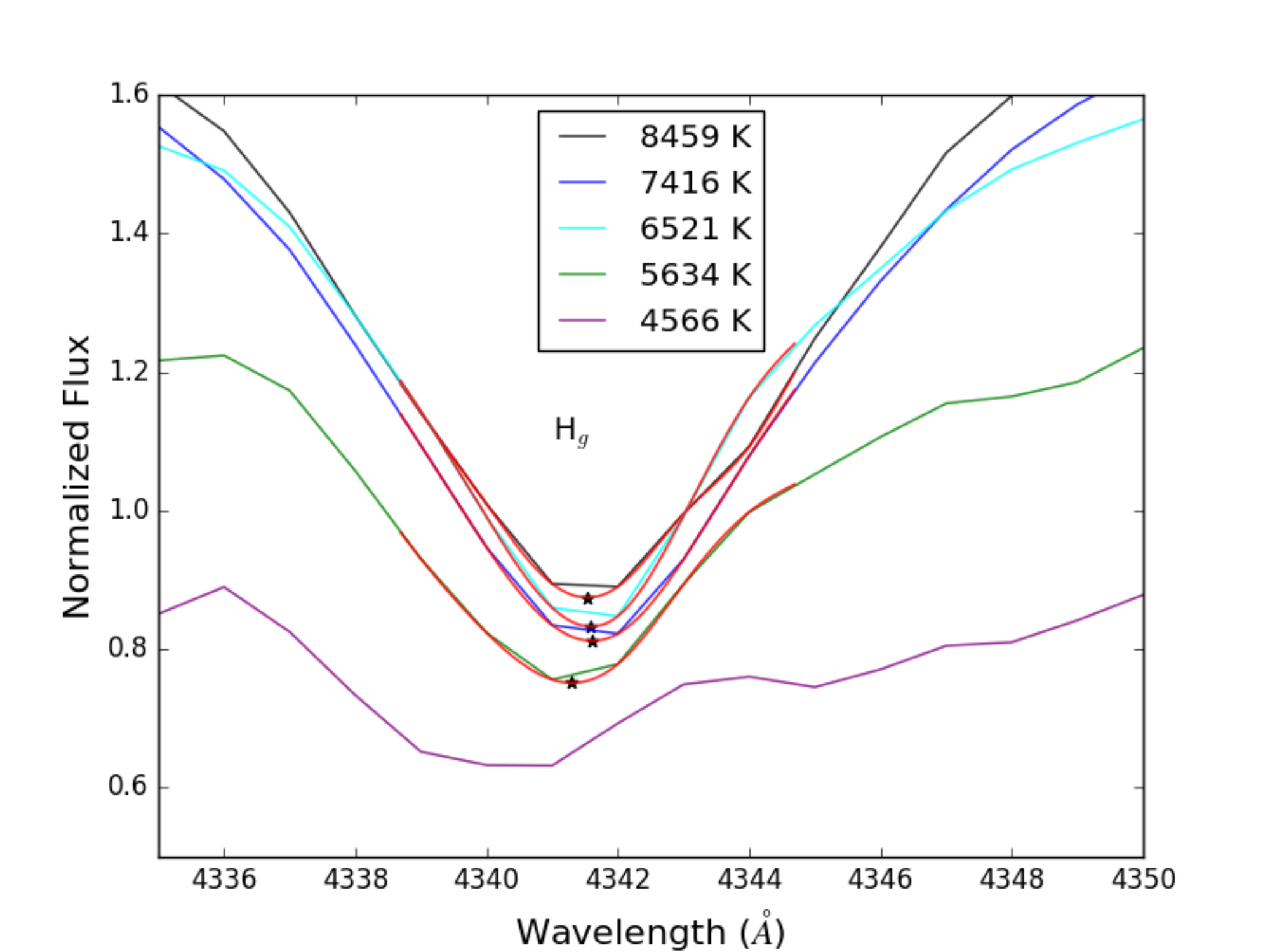}}
\end{minipage}
\vfill
\begin{minipage}{0.5\linewidth}
  \centerline{\includegraphics[width=8cm]{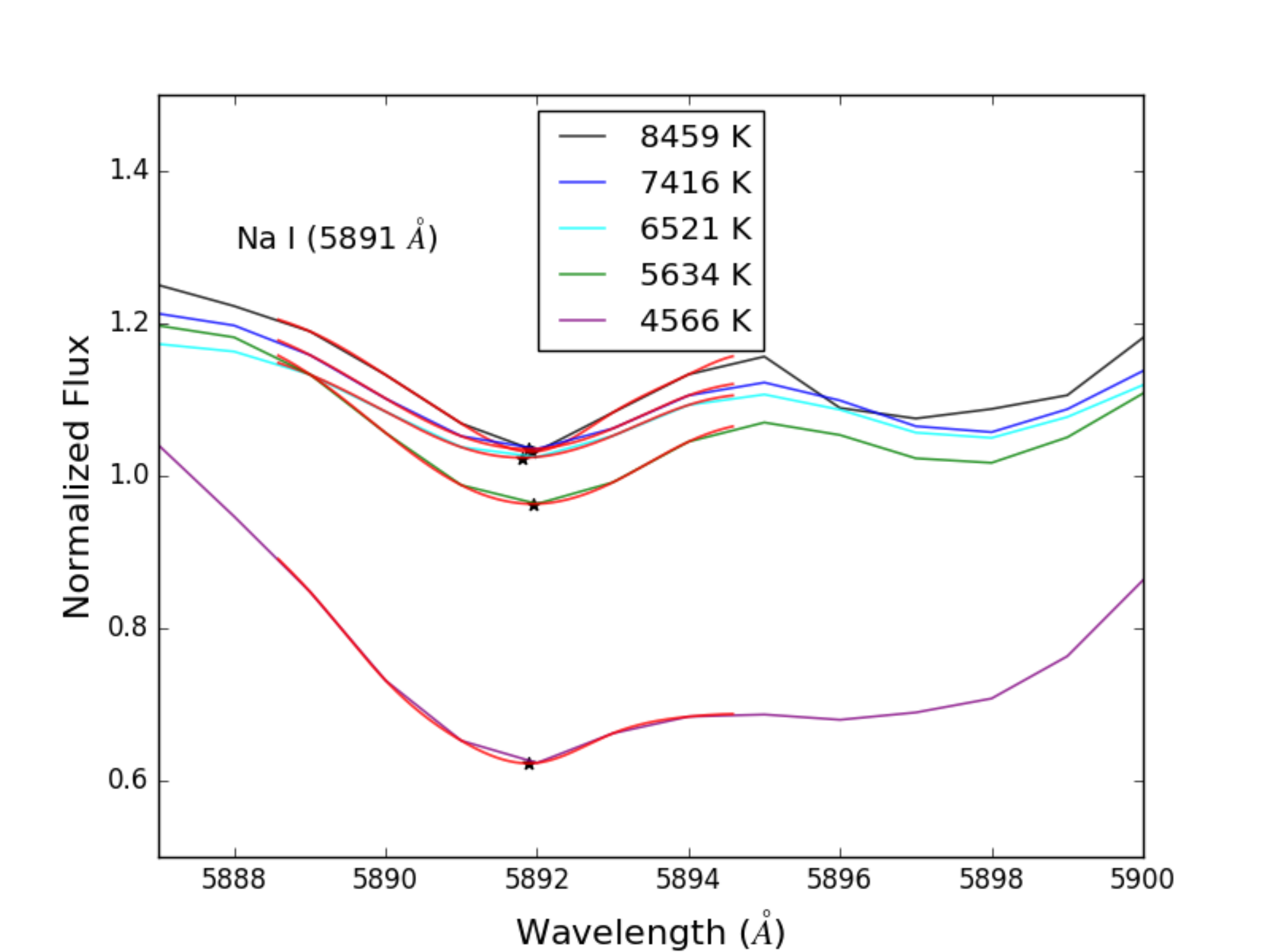}}
\end{minipage}
\begin{minipage}{0.5\linewidth}
  \centerline{\includegraphics[width=8cm]{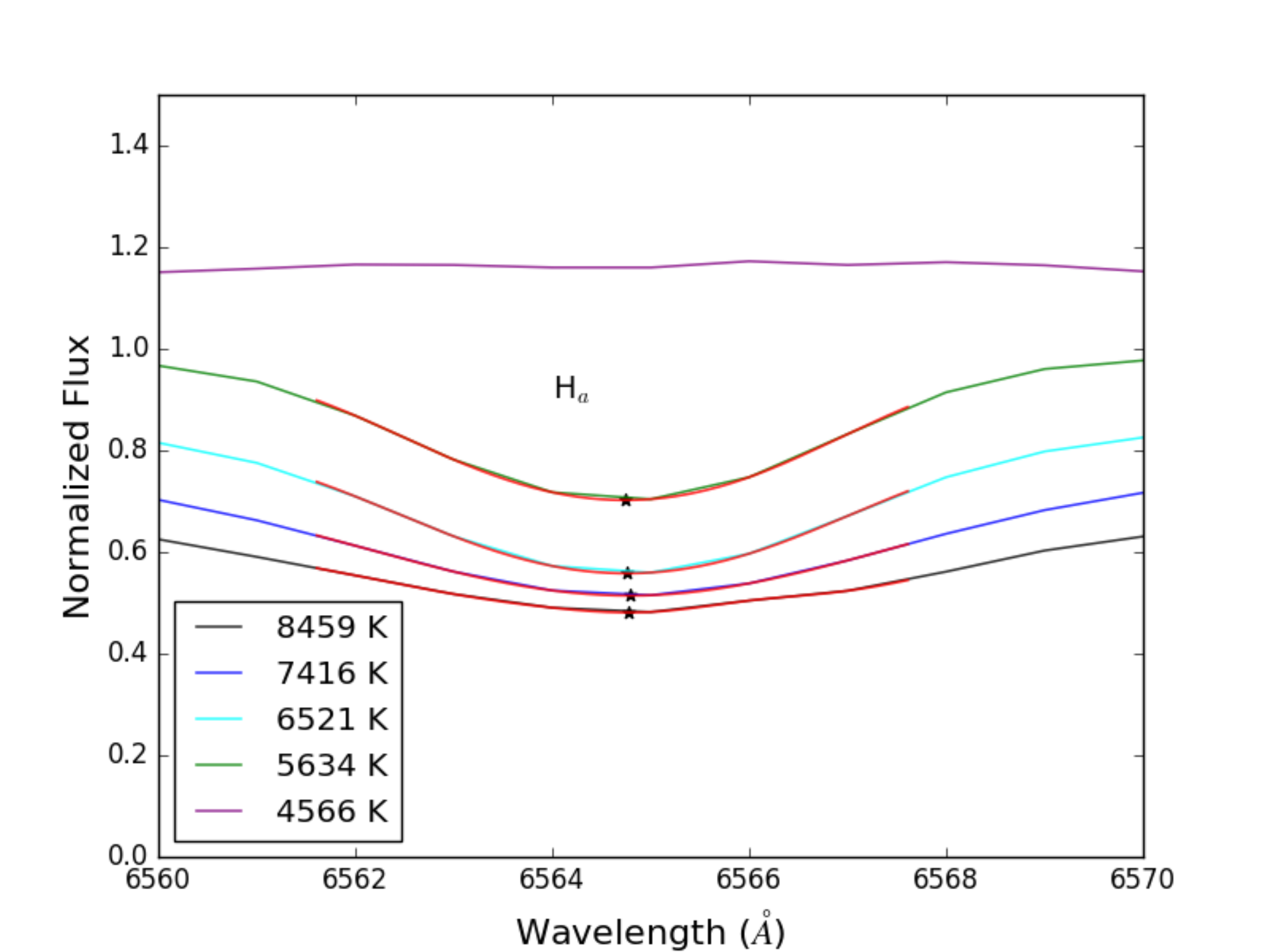}}
\end{minipage}

\caption{Examples of the fitting process of the dense spline interpolations  for Ca II K, Ca II H, H$\gamma$ (upper panel), Na I,  and H$\alpha$ lines (lower panel),  where H$_{g}$ for H$\gamma$ and H$_{a}$ for H$\alpha$.  The smooth red curves are the results of the dense spline interpolations, and the black asterisks are the fitted center wavelengths of the lines. For K type stars, shown in purple, we did not fit Ca II K, Ca II H, H$\gamma$ and H$\alpha$ lines owing to their weaknesses.}
\label{fig12}
\end{figure}

\begin{figure}
\begin{minipage}{0.5\linewidth}
  \centerline{\includegraphics[width=8cm]{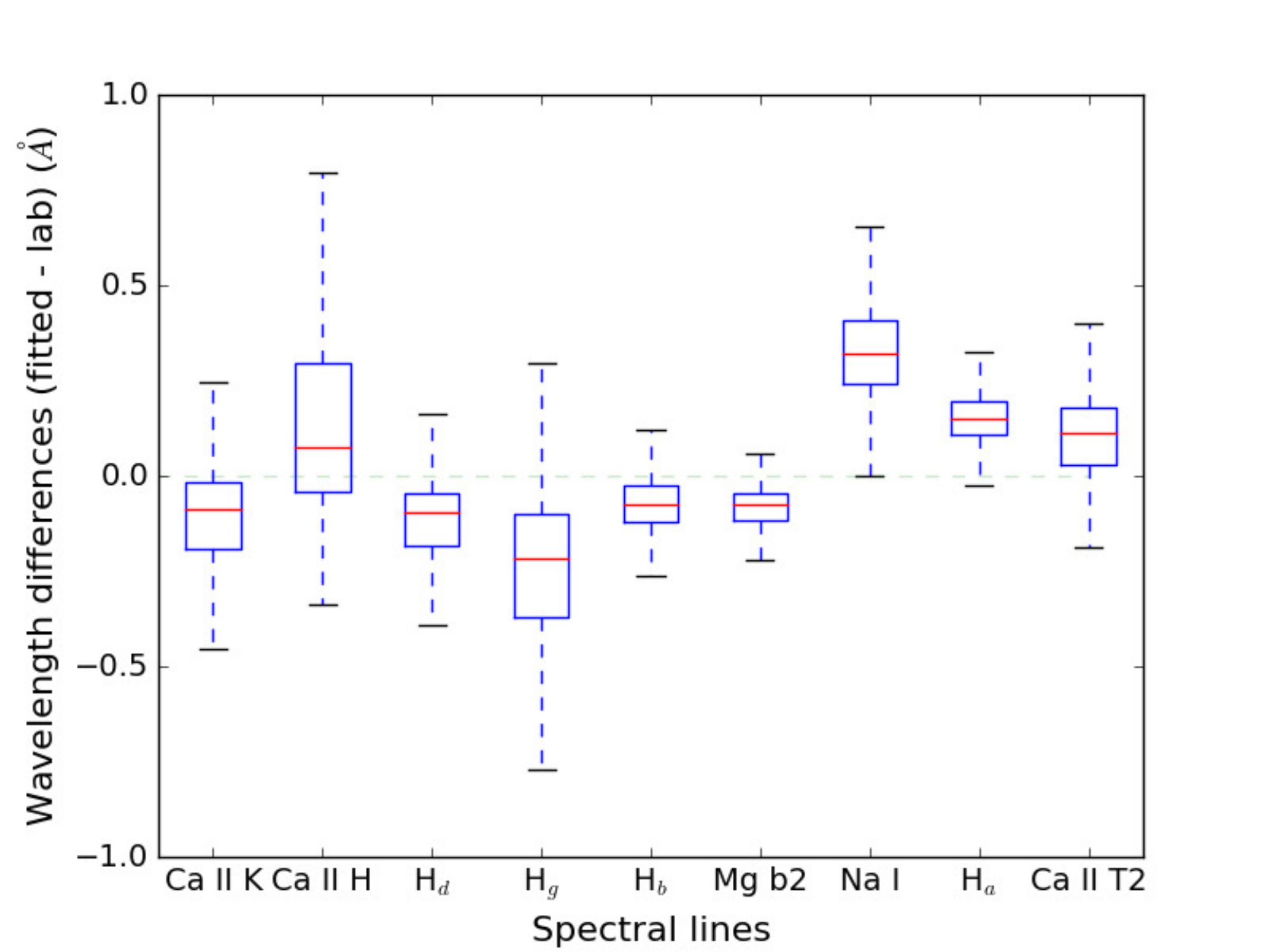}}
\end{minipage}
\begin{minipage}{0.5\linewidth}
  \centerline{\includegraphics[width=8cm]{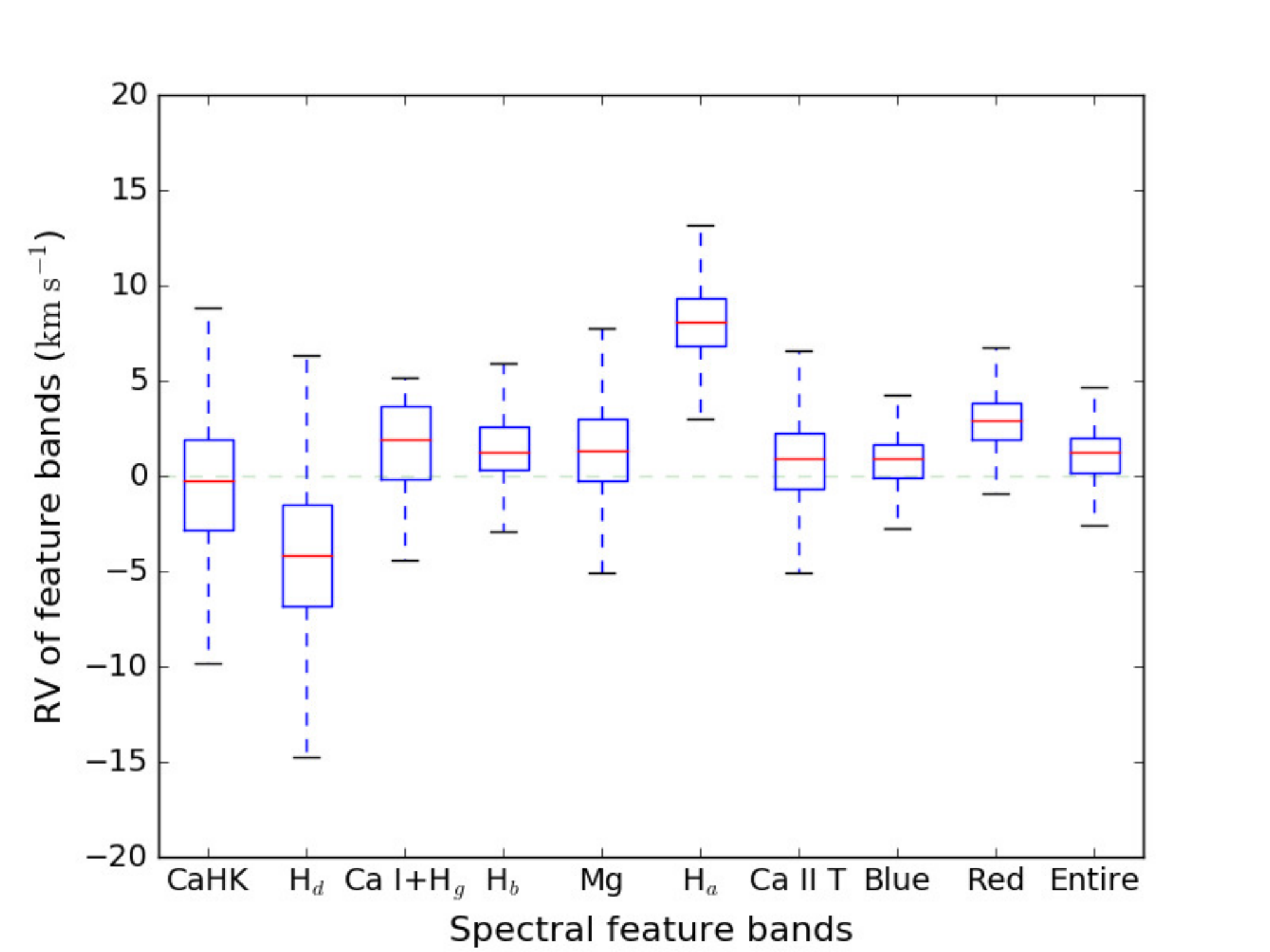}}
\end{minipage}
\caption{Left: distributions of differences between the fitted center wavelength and the laboratory wavelength of the Ca II K, Ca II H,  H$\delta$ (H$_{d}$), H$\gamma$ (H$_{g}$), H$\beta$ (H$_{b}$), Mg b2 (5174.125 \AA ), Na I (5891.583), H$\alpha$ (H$_{a}$) and Ca II T2 (8544.44 \AA ) lines of our library spectra. Right: distributions of the individual velocities calculated from the 10 spectral feature bands of our library spectra. For both, the box extends from the lower to upper quartile values of the differences (left),  and the velocities (right)}, with a red line at the median.
\label{fig13}
\end{figure}

When using an empirical stellar spectra library, some users may consider only certain individual spectral feature lines. Therefore, it is necessary to analyze the wavelength accuracy of individual spectral lines. To verify that the individual spectral lines are accurate in the templates, we fitted nine spectral lines of the 2892 template spectra using a dense spline interpolation method (see Fig \ref{fig12}),  and we obtained their center wavelengths. The observed spectral lines, particularly the blended lines,  may not be a Gaussian or Lorenzian function  of  wavelength. Thus,  we adopted a dense spline interpolation method instead of lines fitting with a Gaussian or Lorenzian function to avoid fitting failure.   The nine spectral lines are Ca II K, Ca II H, H$\delta$, H$\gamma$, H$\beta$, Mg b2 (5174.12 \AA \  vacuum), Na I (5891.58 \AA \ vacuum), H$\alpha$, and Ca II T2 (8544.44 \AA \ vacuum).   For K stars, we ignored weak lines and selected Mg b2 (5174.12 \AA \  vacuum), Na I (5891.58 \AA \ vacuum)  and Ca II T2 (8544.44 \AA \ vacuum).  For G and F stars, we selected all nine spectral lines. For late-type A stars, we ignored  Mg b2 (5174.12 \AA \  vacuum) and Ca II T2 (8544.44 \AA \ vacuum) owing to their weaknesses. We compared the  fitted center wavelengths with the corresponding laboratory wavelengths. Fig \ref{fig13} (left) shows the distributions of differences between the fitted center wavelengths and the laboratory wavelengths. We noticed a small systematic higher shift  for Na I (5891.58 \AA \ vacuum) and   H$\alpha$.  The  Na I (5891.58 \AA \ vacuum) was blended with Na I (5897.55 \AA \ vacuum ),  resulting in an asymmetrical flux distribution (see Fig \ref{fig12}). Therefore,  the fitted center wavelengths may deviate from the true values. However, the cause of the small wavelength shift of  H$\alpha$ line remains unclear.  In general, this magnitude of wavelength uncertainties at the LAMOST spectral resolution of R $\sim$ 1800  verifies that the individual spectral lines are accurate in the rest frames of our templates except H$\alpha$ showing small wavelength shifting.

To test whether the small wavelength shift of H$\alpha$ affects the measurements of radial velocities, we calculated the individual radial velocities from 10 spectral feature bands:  CaHK (3900--4000 \AA), H$\delta$ (4080--4120 \AA), Ca I+H$\gamma$ (4200--4500 \AA), H$\beta$ (4800--4950 \AA), Mg (5100--5250 \AA), H$\alpha$ (6520--6595 \AA), Ca II Triplet (8400--8700 \AA), blue arm(3800--5900 \AA), red arm (5900--8900 \AA),  and  the entire spectrum (3800--8900 \AA). Corresponding to the 10 feature bands, we divided each spectrum of this library into 10 regions, that were then cross-correlated with the reference spectra using the corresponding parameters from the KURUCZ spectral library. For the CaHK, H$\delta$, H$\beta$ and  H$\alpha$ bands, we calculated their individual radial velocities from spectra with $T_{\rm eff } > $ 5000 K,  because these features are weak in cool stars, this results in large uncertainties in velocity estimations from these lines. For the same reason,  when we calculated the velocity of Ca II Triplet, we did not consider the hot stars ($T_{\rm eff} > $ 7500 K).  The distributions of the individual velocities calculated from the 10 spectral feature bands are shown in Fig \ref{fig13} (right). We noticed  a systematic higher shift of $\sim$ 8 ${\rm km \, s}^{-1}$  for the H$\alpha$ band,  which is consistent with the small wavelength shift shown in Fig \ref{fig13} (left).  This may be attributed to mis--calibration of the wavelength. The cause of this shift remains unclear, in  view of the agreement between the velocities of the Ca II Triplet (red arm) and  those of the spectral feature bands of the blue arm.  If only the H$\alpha$ line of the red arm is used to calculate the velocities, an offset of $\sim $  7 ${\rm km \, s}^{-1}$ will occur. In this case, an RV-red value is produced that is 7 ${\rm km \, s}^{-1}$ larger than the RV-blue value, which is consistent with previous research \citep{boeche}.  Fortunately, when cross correlations are used on the entire spectrum, the  effect of the H$\alpha$ wavelength shift on the radial velocity measurement is negligible.  Fig \ref{fig13} (right) shows that the radial velocities on the entire spectrum are consistent with those on the blue arm. If cross correlations are used on the red arm, the RV offset is smaller than that when only the H$\alpha$ spectral line used ($\sim$ 3 ${\rm km \, s}^{-1}$). It should be noted that the systematic wavelength offset of the H$\alpha$ line is given to facilitate users,  who employ this library to measure radial velocities to make specific choices.  We did not consider the Na I D lines because the Na I D lines,  which are located at the junction of the blue and red arms and  are sometimes not credible,  may introduce large uncertainties in the  radial velocity measurement.

\clearpage

\section{Validations}
To verify the reliability of the stellar atmospheric parameters derived from our library, we performed an internal cross--validation and a couple of external comparisons. We used a simplified version of  SpecMatch--Emp \citep{yee} to compute the parameters of individual spectra from this library.  We did not apply a rotational broadening kernel to account for the relative $v \sin i$ between the target and reference stars,  in consideration of the resolution of the template spectra ($R \sim 1800$). The simplified SpecMatch--Emp interpolates between the parameters of the library spectra by synthesizing linear combinations of the five best--matching spectra (the smallest five $\chi^2$).  The entire spectrum (3900--8800 \AA) was used in the template matching. 

\subsection{Internal cross--validation}

\begin{figure}
\begin{minipage}{0.5\linewidth}
  \centerline{\includegraphics[width=10cm]{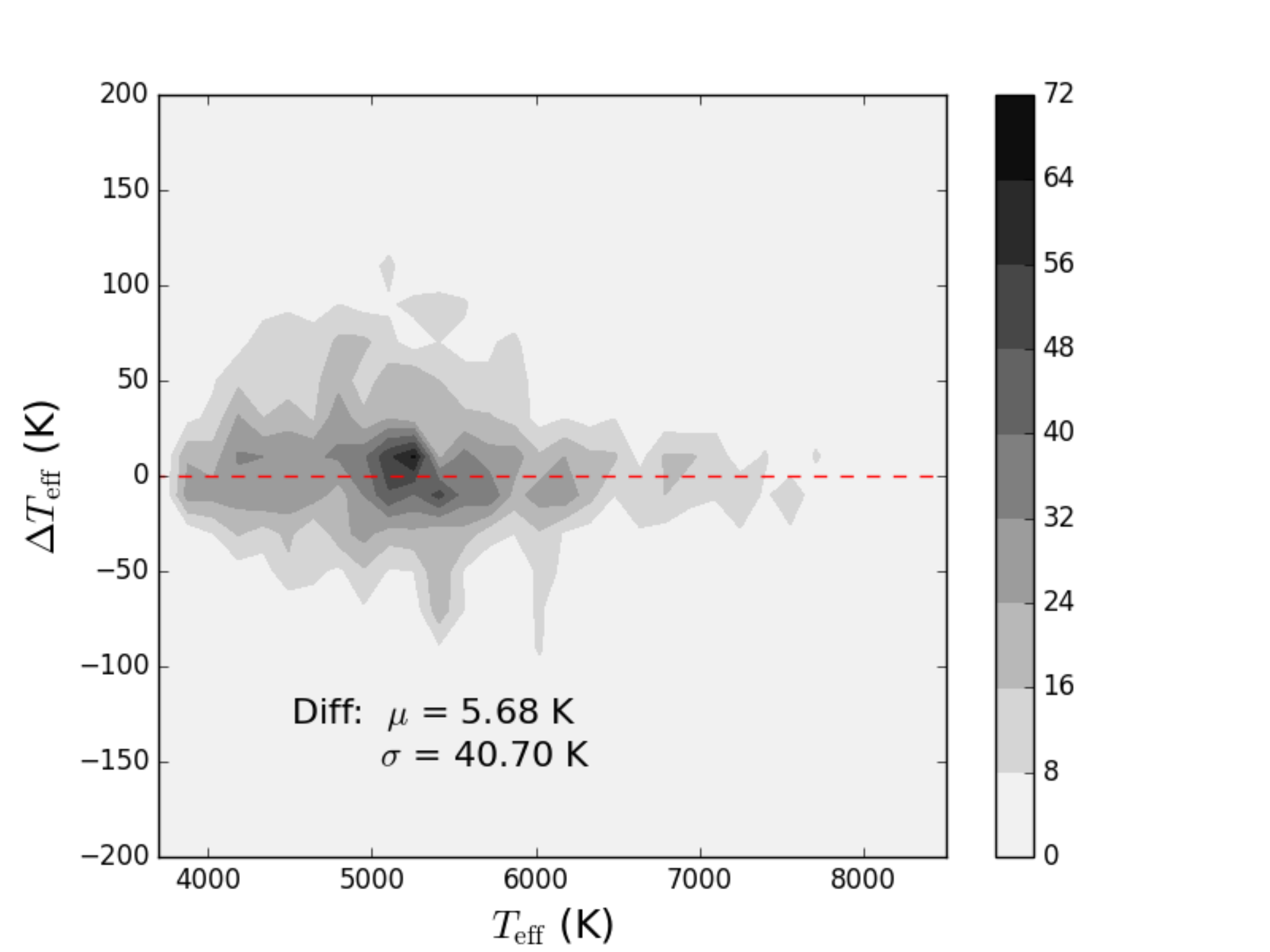}}
\end{minipage}
\begin{minipage}{0.6\linewidth}
  \centerline{\includegraphics[width=10cm]{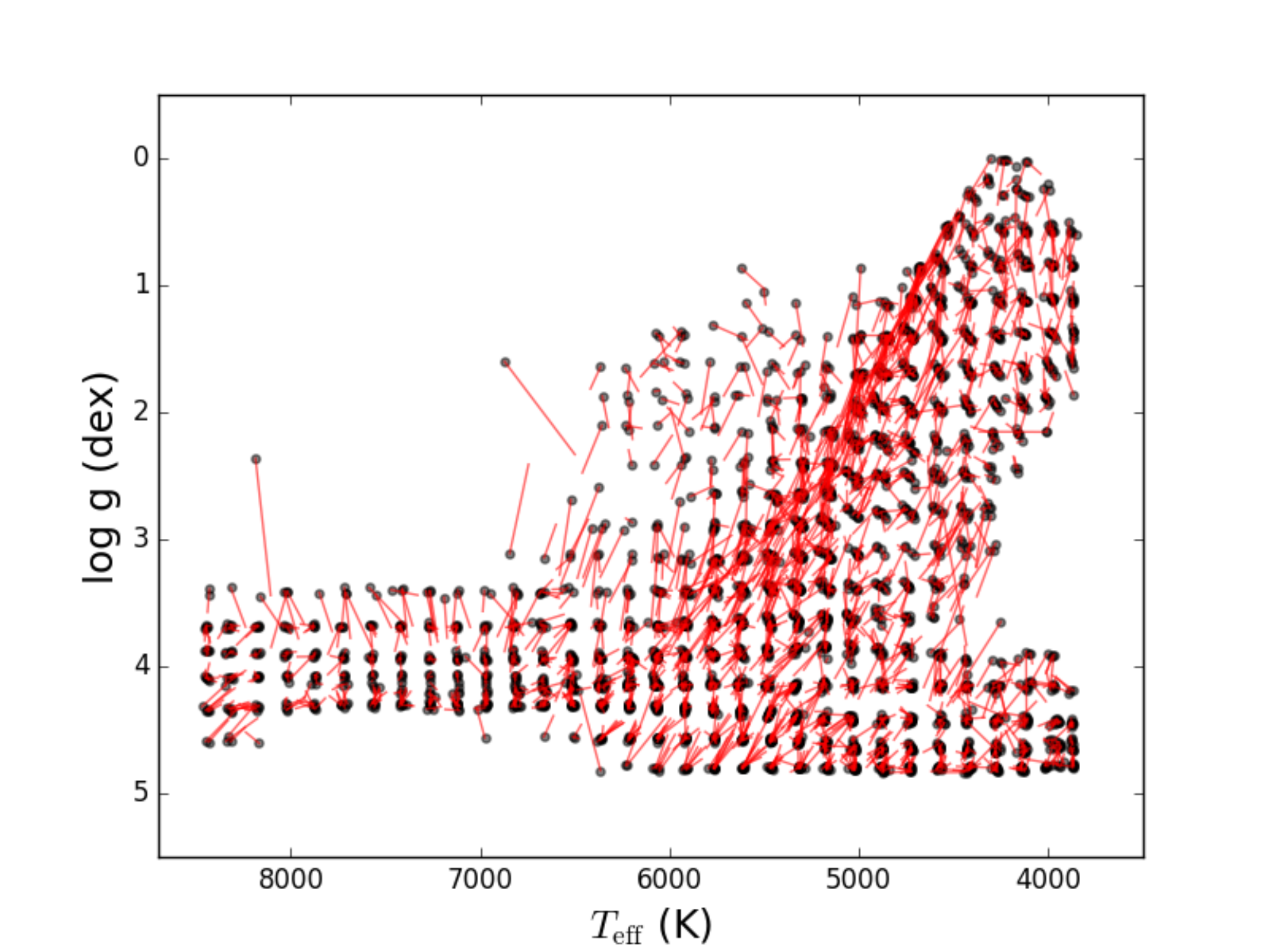}}
\end{minipage}
\vfill
\begin{minipage}{0.5\linewidth}
  \centerline{\includegraphics[width=10cm]{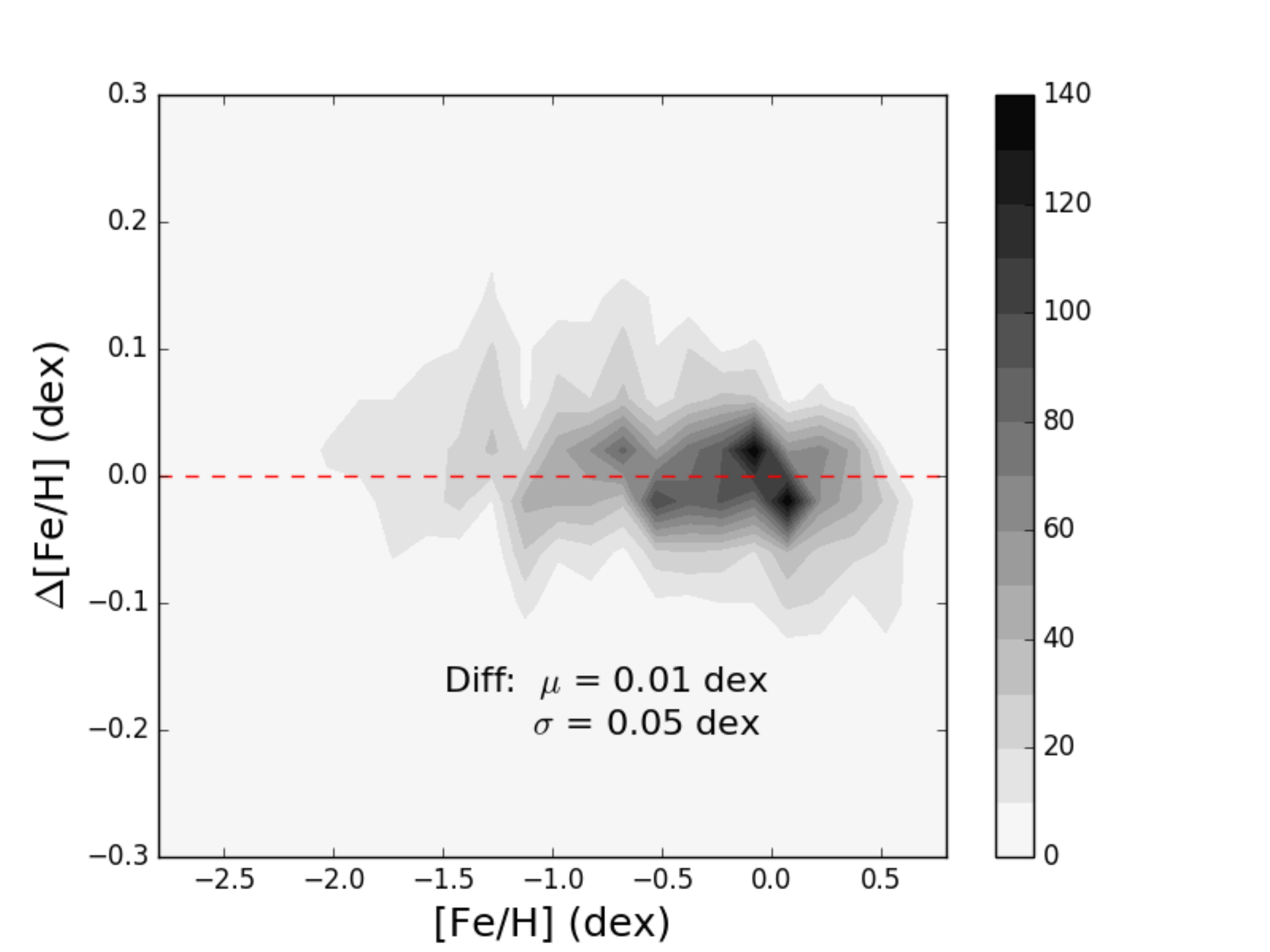}}
\end{minipage}
\begin{minipage}{0.6\linewidth}
  \centerline{\includegraphics[width=10cm]{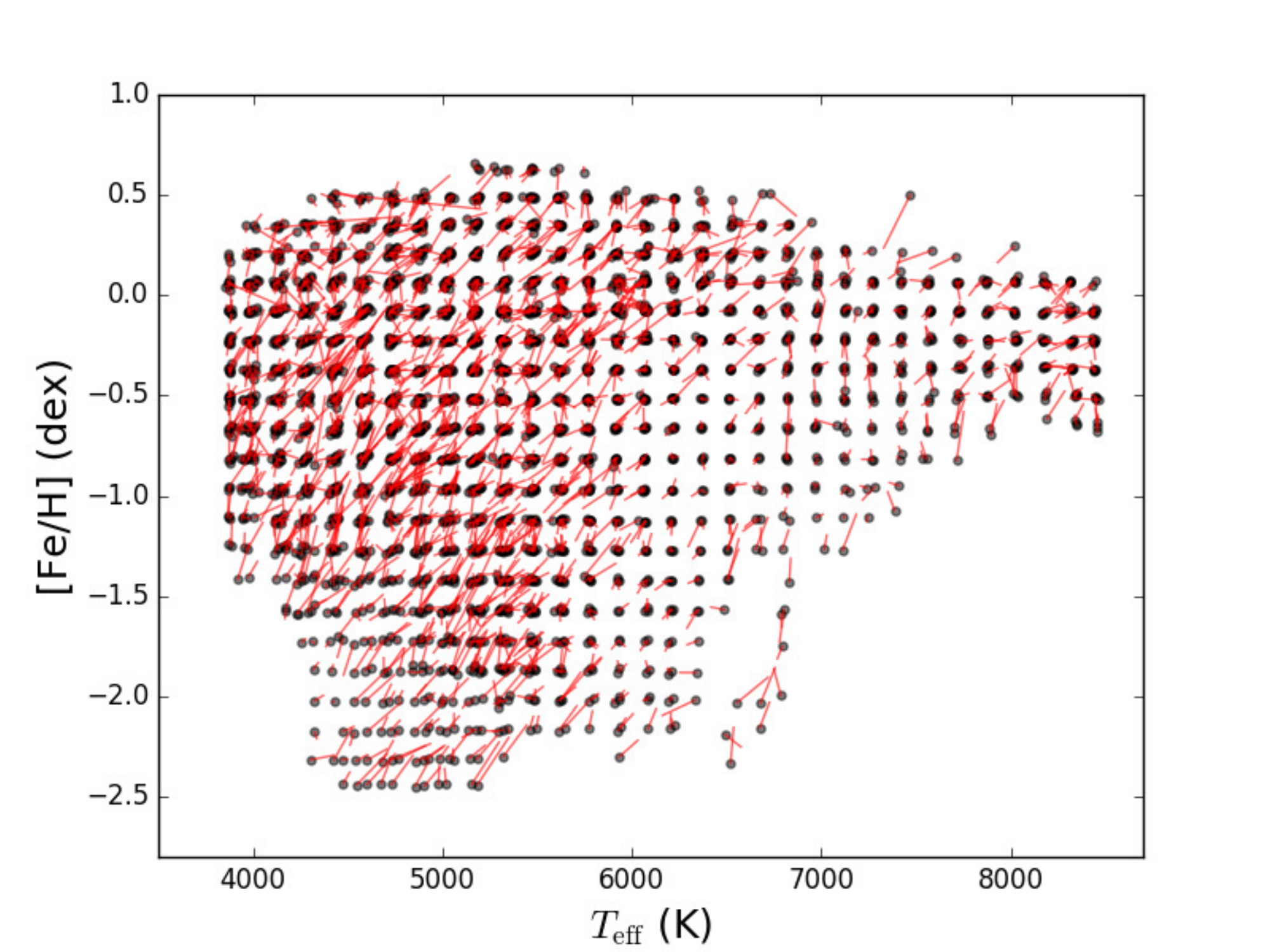}}
\end{minipage}
\vfill
\begin{minipage}{0.5\linewidth}
  \centerline{\includegraphics[width=10cm]{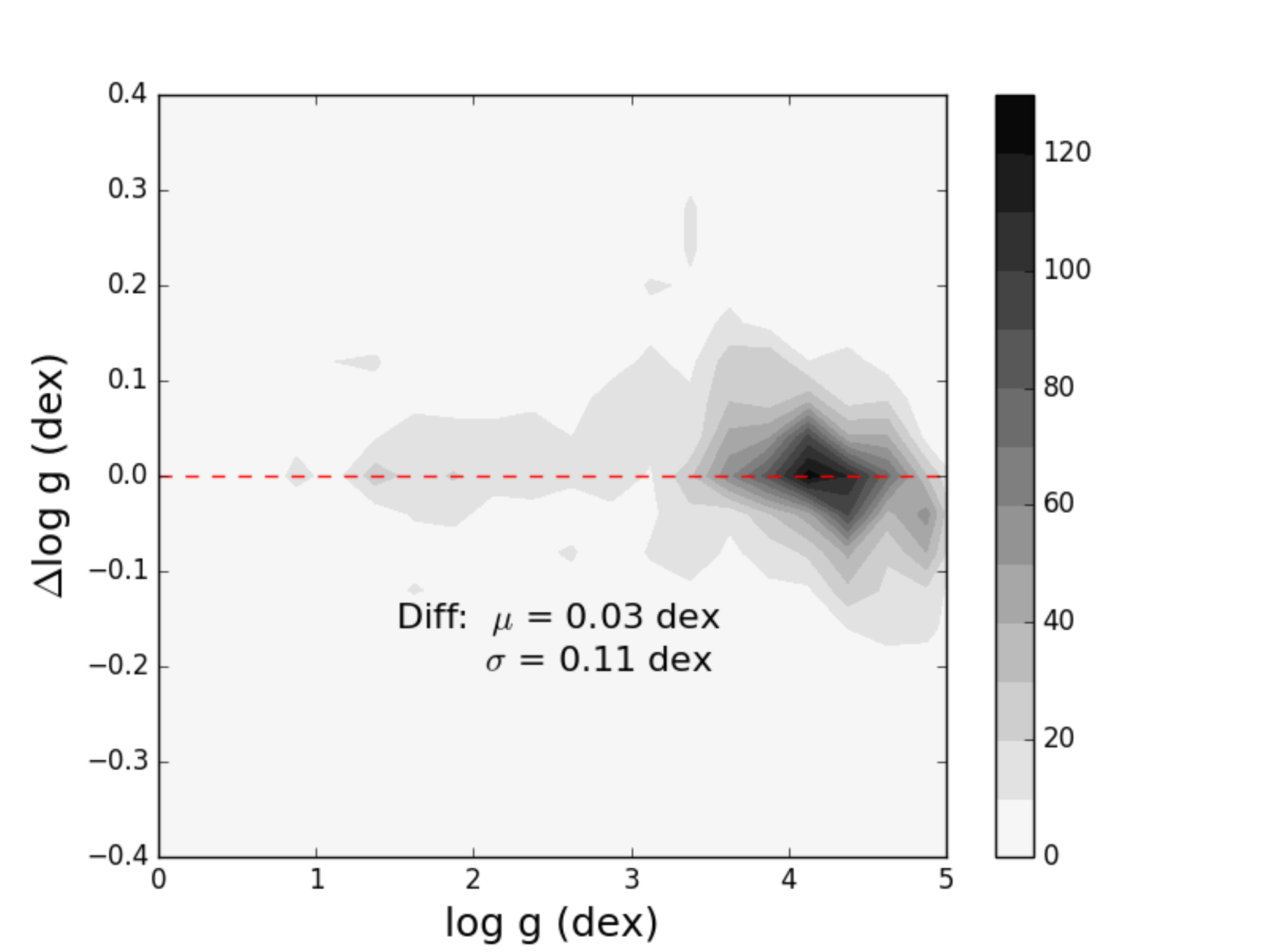}}

\end{minipage}
\begin{minipage}{0.6\linewidth}
  \centerline{\includegraphics[width=10cm]{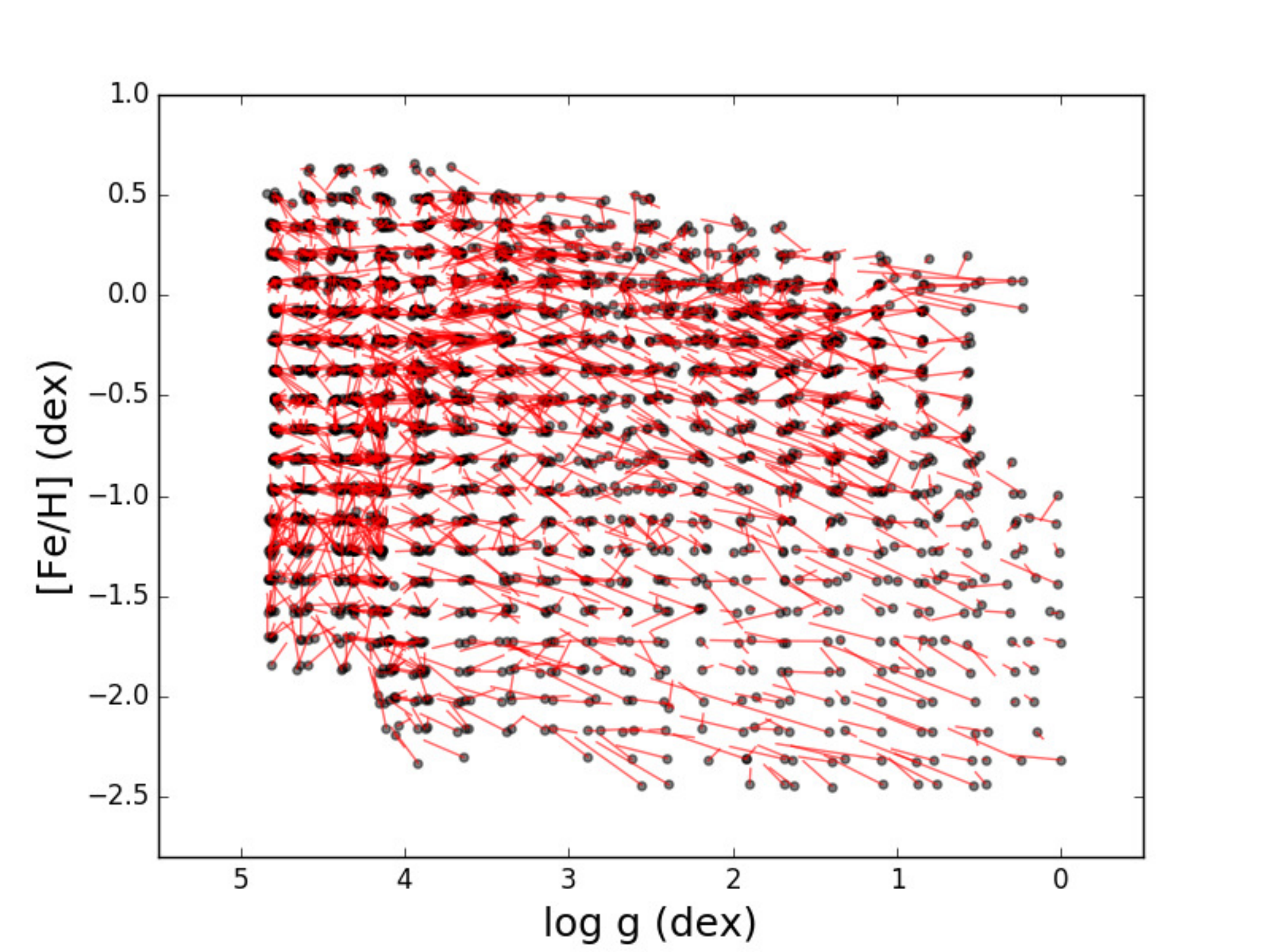}}
\end{minipage}
\caption{ Comparison of library parameters to parameters yielded by the simplified SpecMatch--Emp for each library spectrum in the internal cross--validation process. Left panel: Contour distributions of differences between the library values of $T_{\rm eff}$, [Fe/H], logg and the derived values. Right panel: black points indicate the library stellar parameters, while red lines point to the derived parameters.}
\label{fig14}
\end{figure}

We treated each spectrum in the library as an unknown target and ran the  simplified SpecMatch--Emp to compute its parameters from the remaining library spectra. We then compared these derived parameters to their library values as determined in the previous step. The difference between the derived parameters and the library parameters reflects the errors in the simplified  SpecMatch--Emp and in the library parameters. Fig \ref{fig14} shows the  results of this internal cross--validation. We noted a general tendency for the residuals to be most positive for smaller values of the derived parameters,  and most negative for larger values. This can be partly explained by the fact that our library does not cover an infinite parameter space but occupies only a finite region of parameter space.  Target stars must necessarily match stars in the interior of that parameter space, resulting in their derived parameters being pulled toward the interior of the parameter distribution ( see right panel of  Fig \ref{fig14}).  For all of our library templates, the differences between the library values and the derived values have  a scatter of 40 K in  $T_{\rm eff}$, 0.05 dex in [Fe/H], and  0.11 dex in log \emph{g}. These values are the uncertainties that should be adopted for the output of parameter measurements based on this library.

\subsection{External comparisons}
To verify the reliability of the stellar atmospheric parameters derived from our library, we compared them against stars with stellar parameters in other catalogs. For these comparisons we selected LAMOST spectra with sufficient SNR to allow for a fair estimation of the stellar parameters. 

\subsubsection{Comparison with APOGEE}
\begin{figure}
\centering
\includegraphics[width=14.0cm,height=16.0cm]{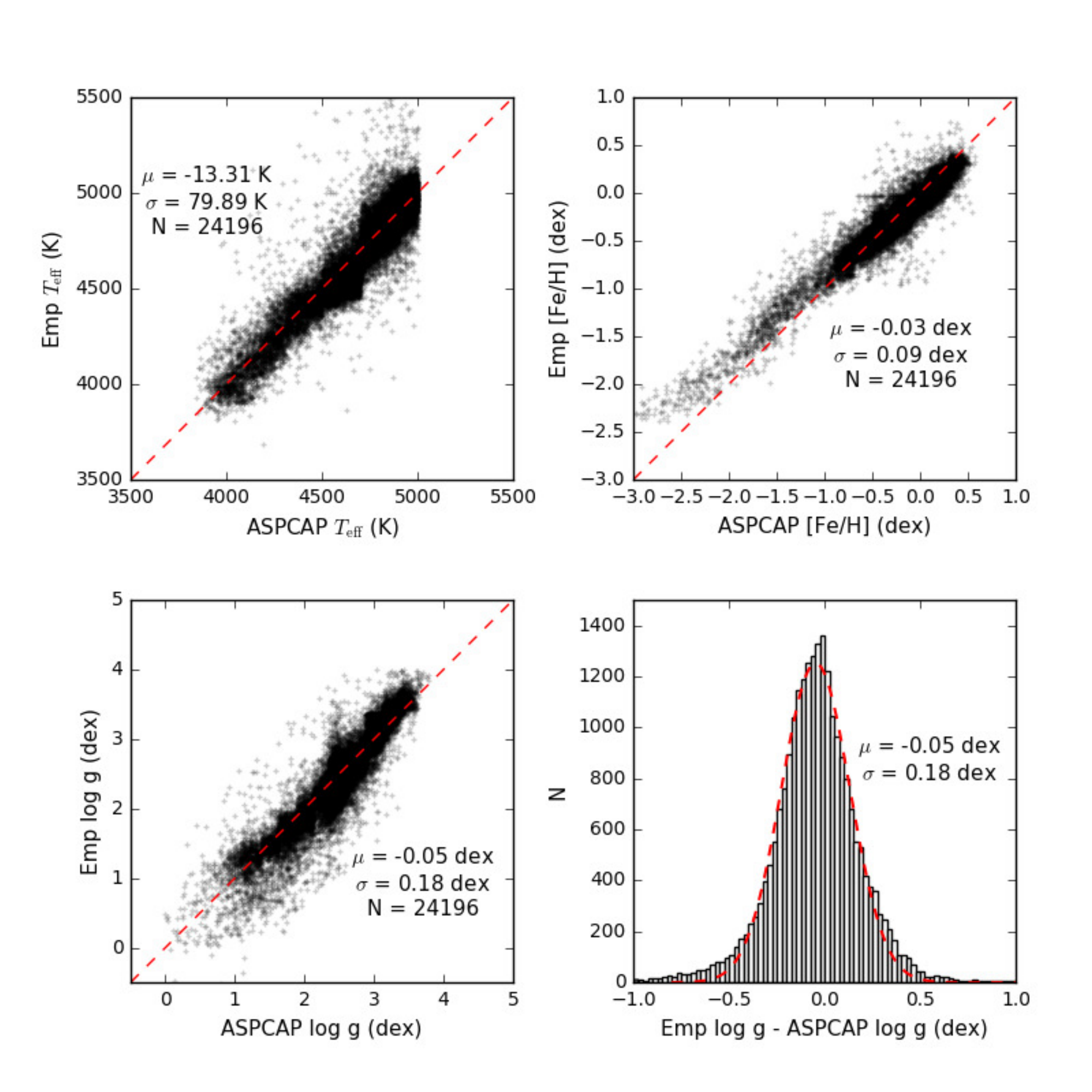}
\caption{ Comparison of the stellar parameters estimated from the LAMOST spectra using our library  to the reference ASPCAP parameters from the APOGEE spectra. The lower right shows the differences between the Emp--log \emph{g} and the ASPCAP--log \emph{g};  no systematic shift was noticed.}
\label{fig15}
\end{figure}

 We cross--matched  LAMOST DR5 with the APOGEE of  the SDSS DR14 and obtained  24, 196 spectra  of common targets with  LAMOST SNRg $>$10.0 (including repeat observations).   
The simplified SpecMatch--Emp using our library, hereafter referred to as Emp,   determined the atmospheric parameters for all 24, 196 LAMOST spectra. The comparison of the two catalogs is shown in Fig \ref{fig15}. The results from this library show very good consistency with the APOGEE parameters. The differences between them have  a scatter of 80 K in  $T_{\rm eff}$, 0.09 dex in [Fe/H], and  0.18 dex in log \emph{g}.  We noted that for APOGEE [Fe/H] $<$ -1.5 dex, metallicities from this library  are systematically higher than those in APOGEE.  The same trend appears in the metallicity comparison between LASP and APOGEE. This metallicity offset may be related to the abundance coverage of our library and the ELODIE library.  Both lack  adequate coverage of low metallicity stars.  The matches in the interior of the parameter space result in the low metallicities derived from them being pulled toward richer metallicities.   For log \emph{g}, we found that the surface gravity offsets between LASP and ASPCAP  were corrected through our library.

\subsubsection{Comparison with Kepler K Giants}
\begin{figure}
\centering
\includegraphics[width=14.0cm,height=16.0cm]{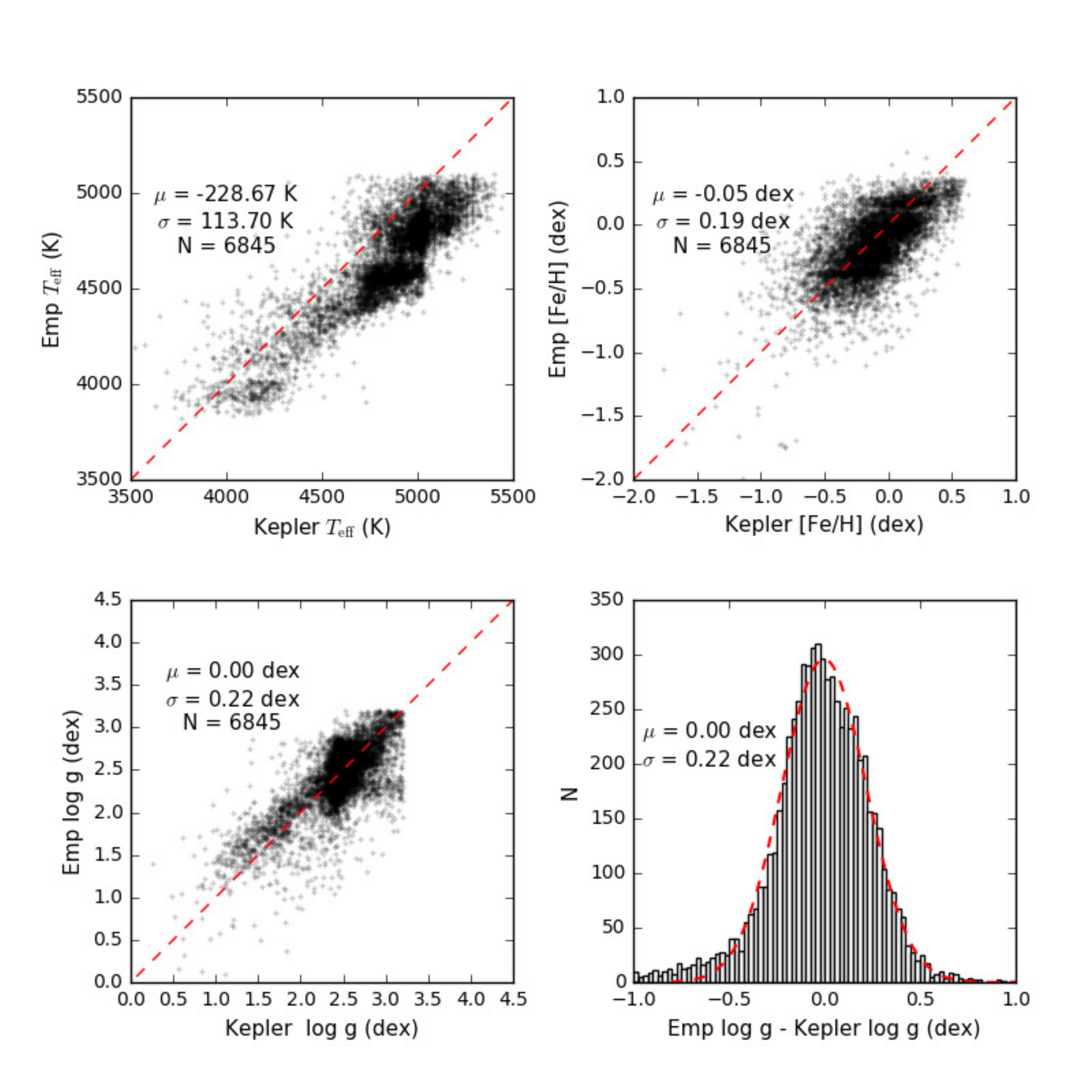}
\caption{ Comparison of the stellar parameters  estimated from the LAMOST spectra using our library to the reference Huber parameters from the NASA Kepler mission. The lower right shows the differences between the Emp--log \emph{g} and the Kepler--log \emph{g};  no systematic shift was noted. }
\label{fig16}
\end{figure}

The properties of about 20, 000 stars observed by the National Aeronautics and Space Administration (NASA) Kepler mission have been reported \citep{huber}. Most of the  stellar parameters  of these stars were acquired from a collection of different catalogs with different observational techniques.  About 15,500 of these stars have precise measurements of surface gravity obtained through asteroseismology,  and can therefore be used as a reference for surface gravity for comparison purposes. The LAMOST DR5 has 6845 stars in common with Huber et al. (2014) with  LAMOST SNRg $>$ 10.0.  These stars are K giants with $T_{\rm eff} < 5100 K$ and log \emph{g} $<$ 3.2 dex. We determined the stellar parameters for the 6845 stars from LAMOST spectra by using our templates.   Fig \ref{fig16} shows a comparison of differences between the parameters measured from our library and the Huber parameters. An offset was apparent in $T_{\rm eff} $, and a poor match in [Fe/H] was noted between the two catalogs.  A systematic overestimation in  Huber' s $T_{\rm eff} $  has been reported by Huber et al.(2014, top panel of Fig 7). The temperature offset in Fig \ref{fig16} appears to be consistent with that found by Huber et al.(2014). This offset is attributed to the limitations of the Kepler Input Catalogue (KIC), from which the values were obtained. The poor match of  metallicities  is not important because the Huber metallicities, which are derived mainly from the KIC, are known to have poor precision \citep{boeche}. The values of Huber' s  log \emph{g} are widely accepted to be extremely accurate ($\sim$ 0.03 dex). The lower panel of Fig \ref{fig16} shows good match with the log \emph{g} derived from our library, with a difference scatter of  0.22 dex.  This difference is still less than the log \emph{g} step (0.25 dex) of our template grid.   The systematic offsets between the spectroscopic surface gravities and the asteroseismic surface gravities were well--corrected through our library.

\subsubsection{Comparison with LASP}
\begin{figure}
\begin{minipage}{0.3\linewidth}
  \centerline{\includegraphics[width=6cm]{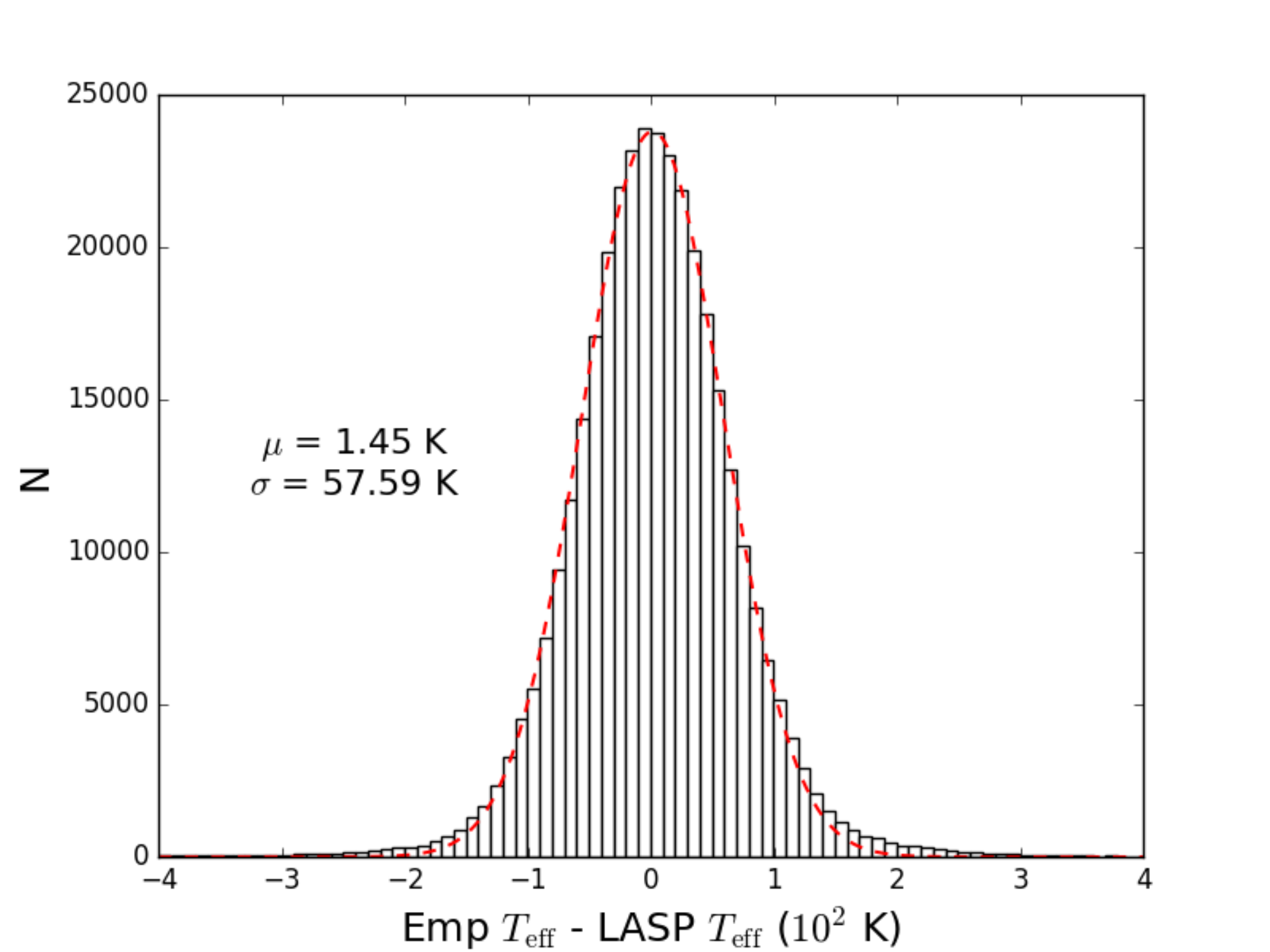}}
\end{minipage}
\begin{minipage}{0.3\linewidth}
  \centerline{\includegraphics[width=6cm]{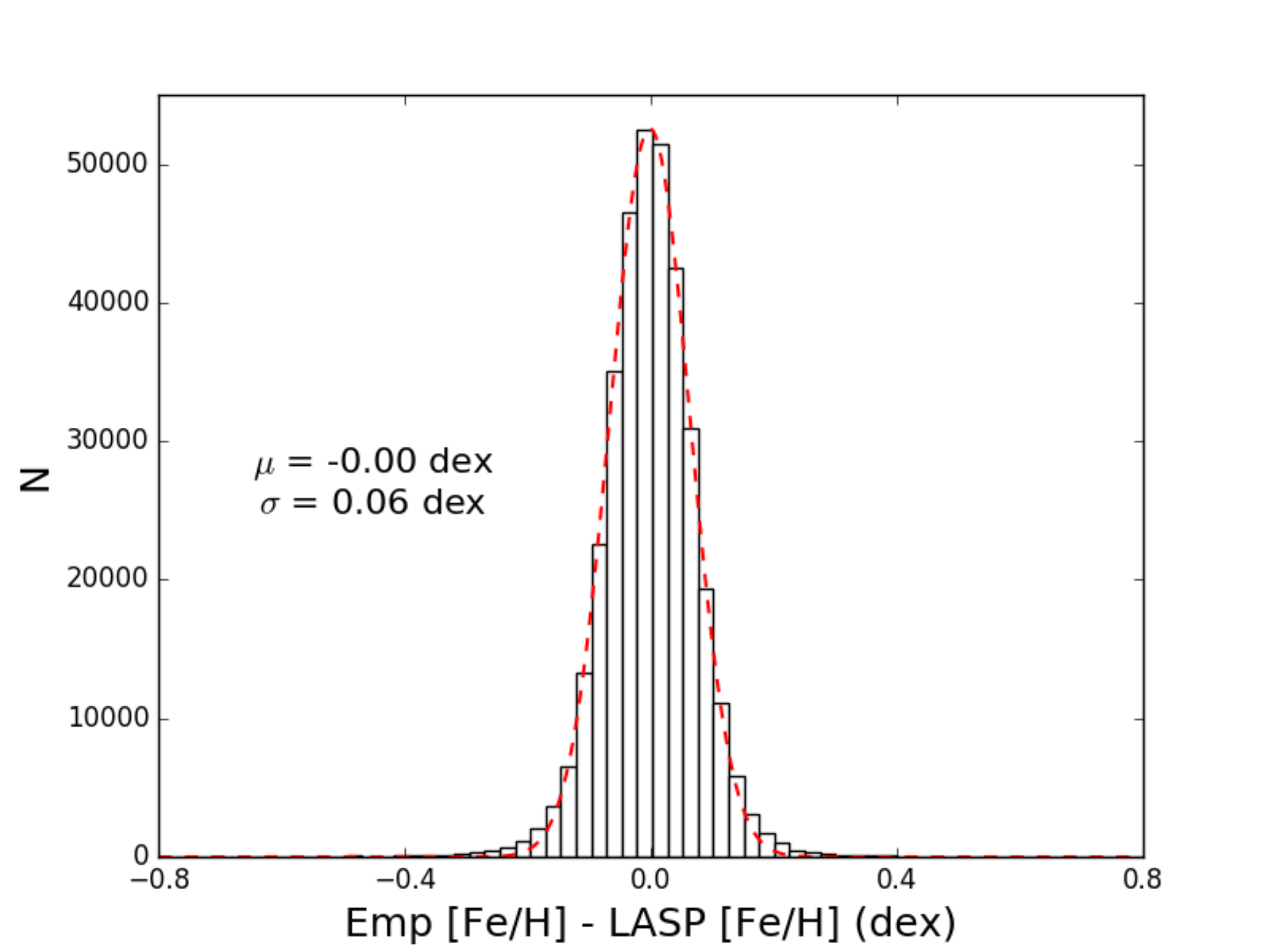}}
\end{minipage}
\begin{minipage}{0.3\linewidth}
  \centerline{\includegraphics[width=6cm]{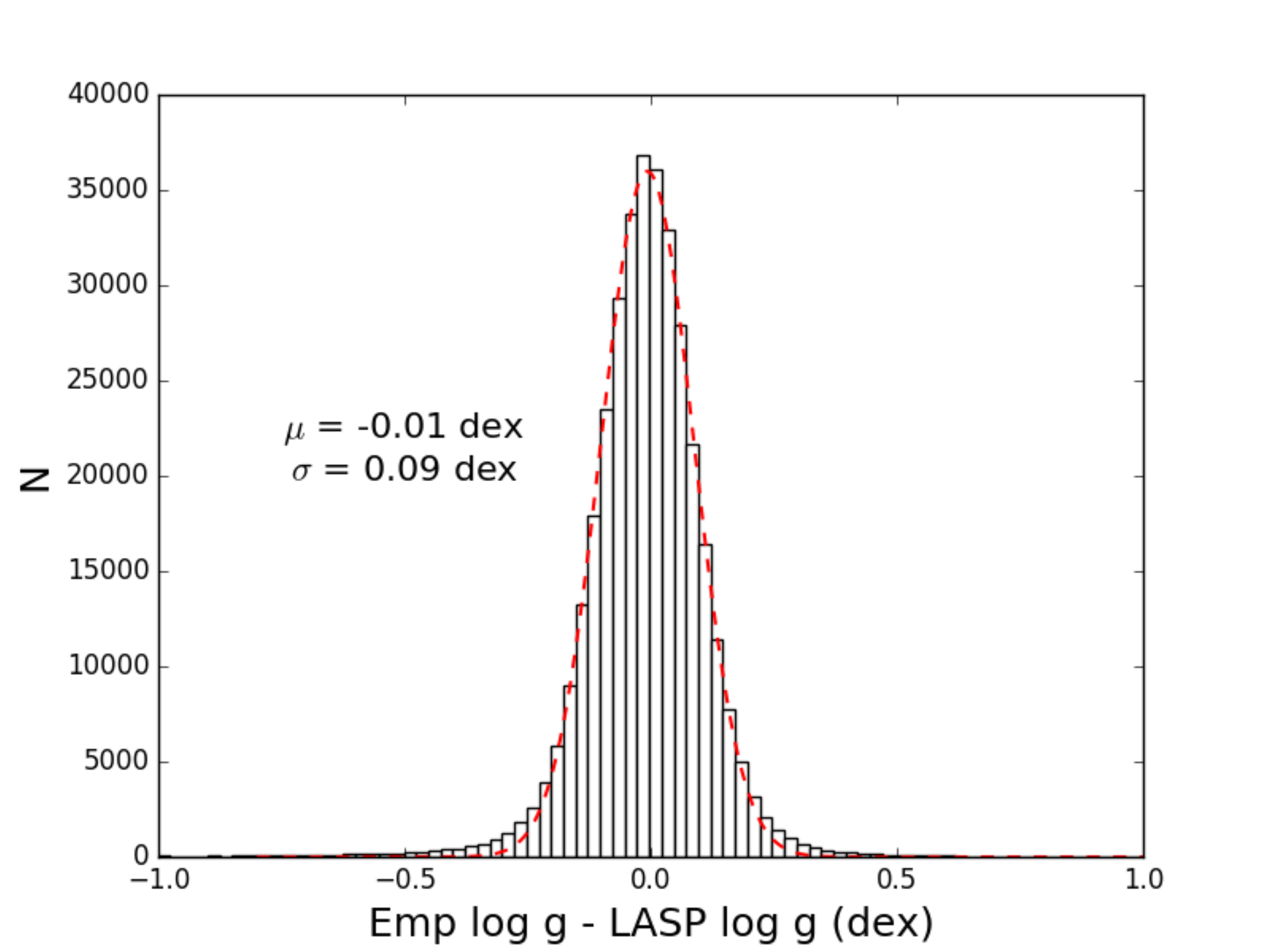}}
\end{minipage}
\caption{Histograms of differences between the parameters measured using our library and LASP. The red dashed curves are Gaussian fits to the distributions,  and the mean and dispersion of the Gaussian fit to the mean and standard deviation values of the differences are also labeled. }
\label{fig17}
\end{figure}

For the main sequence stars of F, G, K and late--type A,  the stellar parameters were obtained directly from LASP. The agreement  between the Emp--derived  parameters and the LASP--derived parameters for these star types validated the parameter labels of our templates. We selected 360,000 spectra of main sequence stars, including 60,000 for late--type A and 100,000 for F,G and K,  with SNRg $>$ 10.0 from the LAMOST database.  We used our library to estimate the stellar parameters of the 360,000 spectra. Fig \ref{fig17} shows the distributions of the differences between the parameters measured by using our library and LASP. The results of  measurement from our library  were highly consistent with those of LASP. This indicates that the parameter labels of our main sequence templates are reliable.

\subsubsection{Comparison with PASTEL} 

\begin{figure}
\begin{minipage}{0.3\linewidth}
  \centerline{\includegraphics[width=6cm]{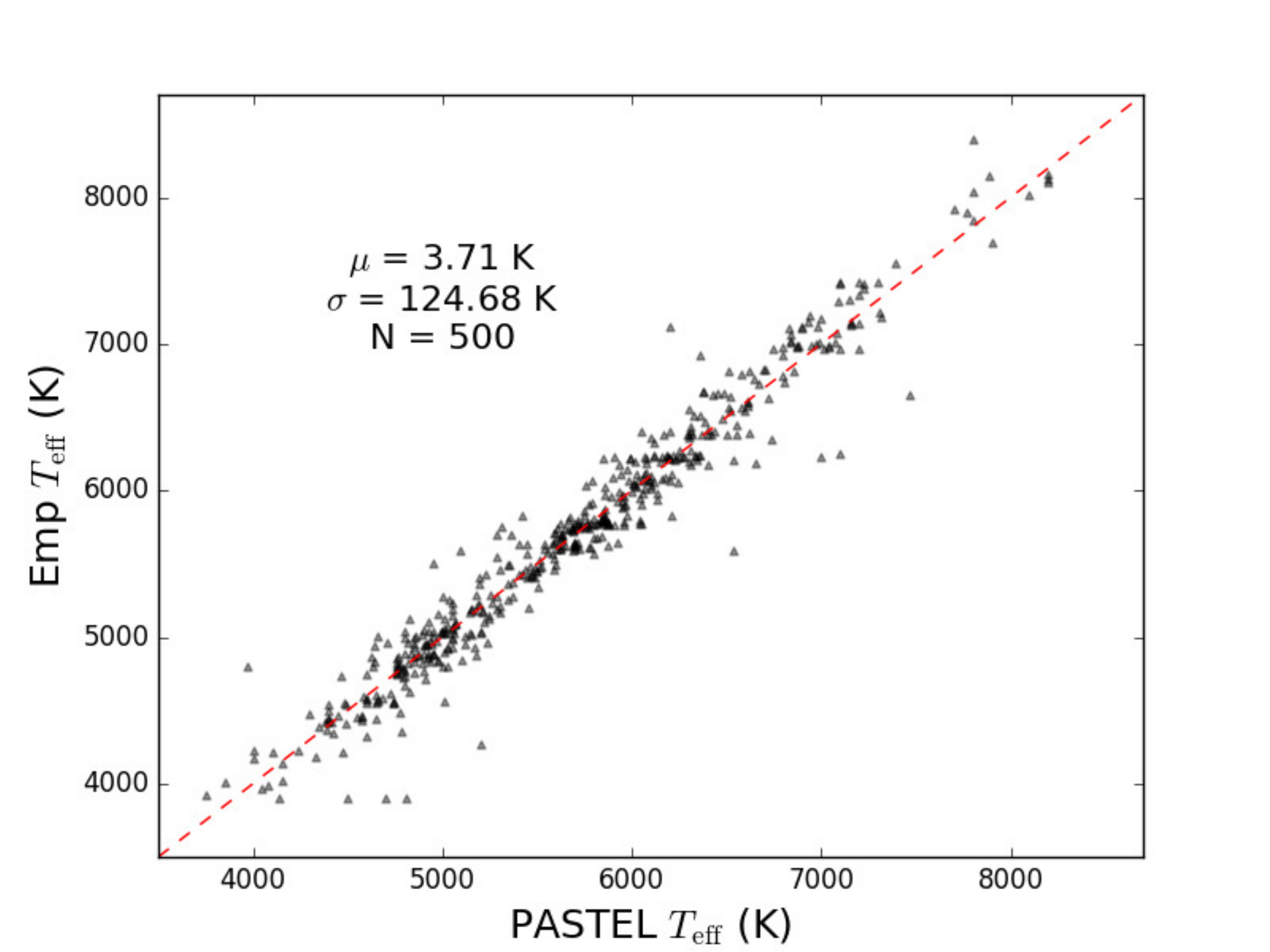}}
\end{minipage}
\begin{minipage}{0.3\linewidth}
  \centerline{\includegraphics[width=6cm]{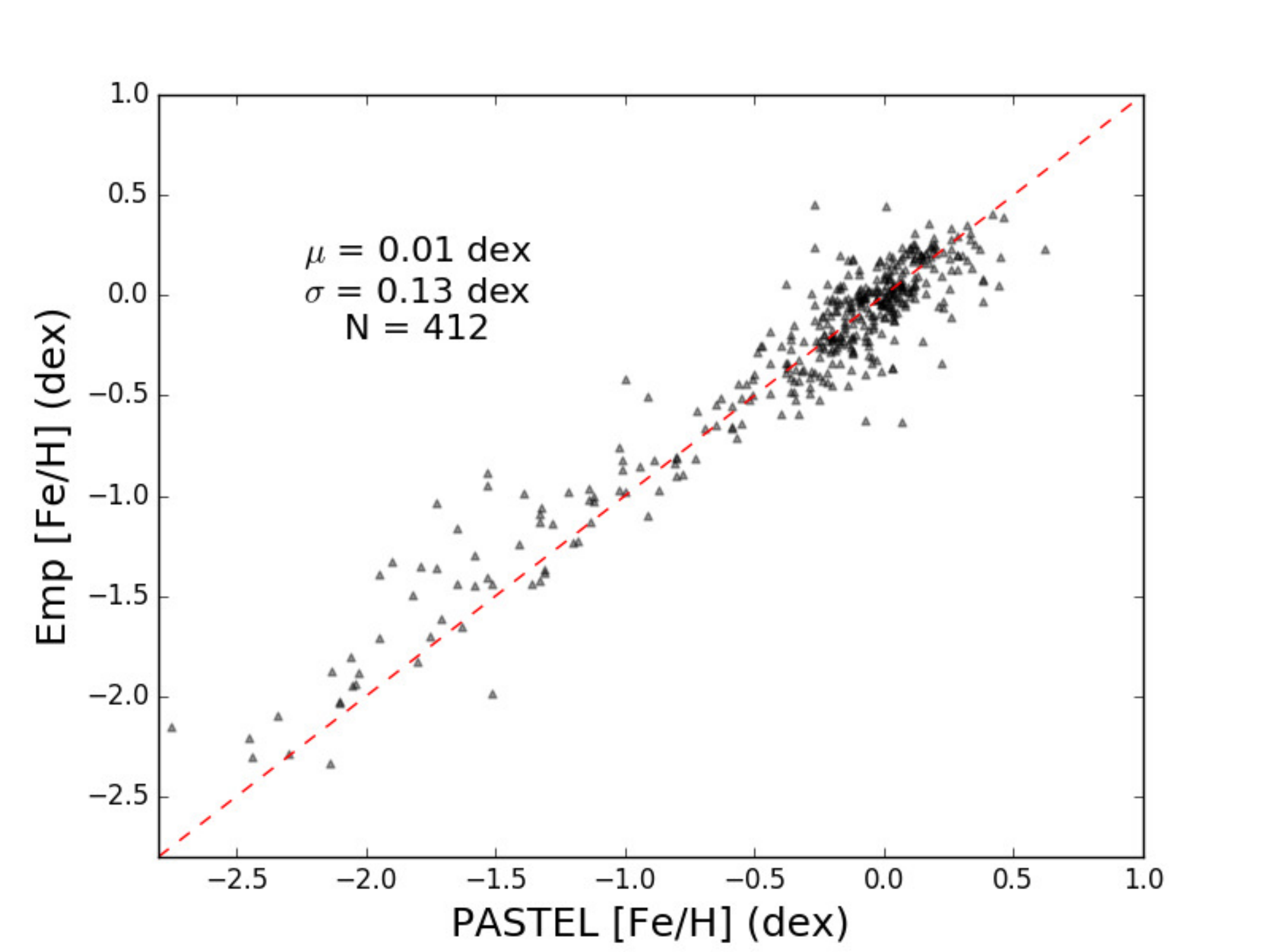}}
\end{minipage}
\begin{minipage}{0.3\linewidth}
  \centerline{\includegraphics[width=6cm]{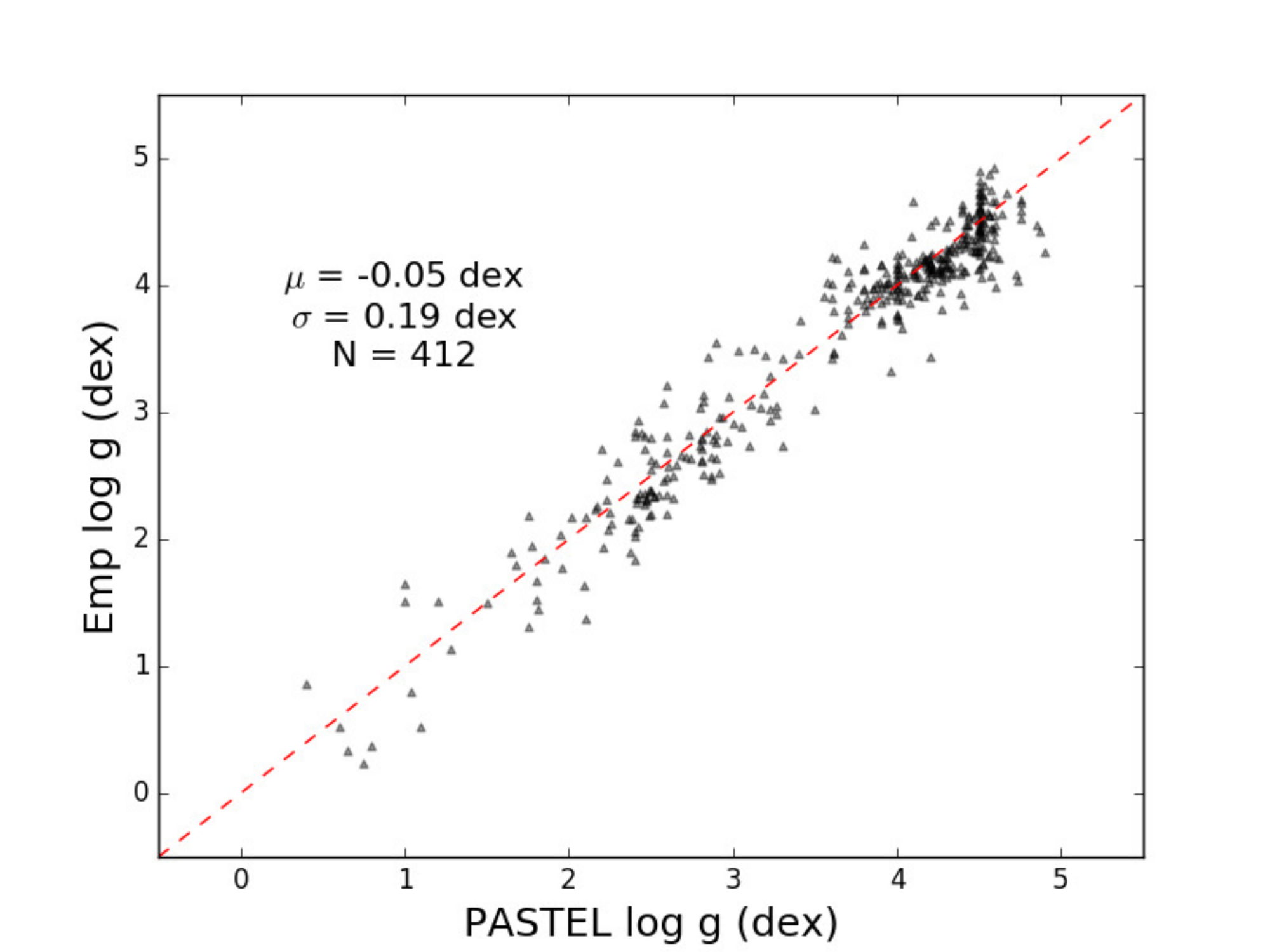}}
\end{minipage}
\caption{ Comparison of the stellar parameters estimated from the LAMOST spectra using our library to the reference PASTEL parameters. }
\label{fig18}
\end{figure}

 The PASTEL catalog includes 30,151 determinations of either $T_{\rm eff }$ or ( $T_{\rm eff }$, log \emph{g}, [Fe/H]) for 16, 649 different stars,  corresponding to 866 bibliographical references \citep{pastel}. Nearly 6000 stars have a determination of the three parameters with high--quality spectroscopic metallicity. We cross--matched LAMOST DR5 with the PASTEL catalog and obtained 500 common stars with  LAMOST SNRg $>$ 10.0, for which 412 stars have a determination of the three parameters ($T_{\rm eff }$, log \emph{g}, [Fe/H]) and the other 88 stars only have temperatures. We determined the stellar parameters of the 500 stars from the LAMOST spectra using our library. Fig \ref{fig18} shows the comparison of  the Emp--derived parameters from  the LAMOST spectra to the reference PASTEL parameters.  We found that the parameters obtained from our library matched the PASTEL parameters fairly well, with small offsets of 125 K for $T_{\rm eff }$, 0.13 dex for [Fe/H], and 0.19 dex for log \emph{g}. For both giant and dwarf stars, no clear systematic offset was found. However, for [Fe/H] $<$ -1.5 dex, a systematic overestimation of metallicities was perceptible owing to the limited number of low metallicity stars.

\subsubsection{Uncertainties and Systematic Errors}

\begin{figure}
\begin{minipage}{0.3\linewidth}
  \centerline{\includegraphics[width=6cm]{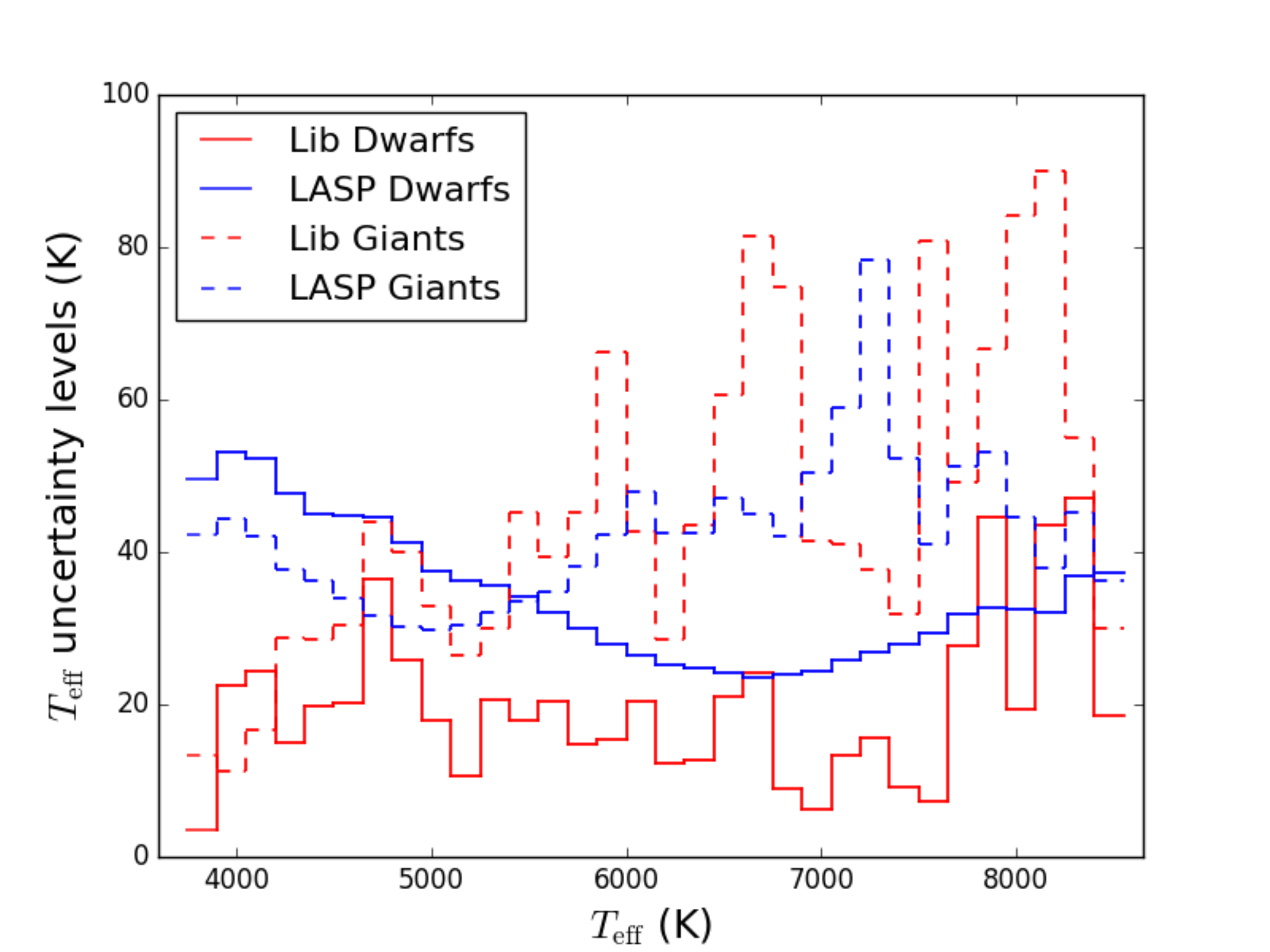}}
\end{minipage}
\begin{minipage}{0.3\linewidth}
  \centerline{\includegraphics[width=6cm]{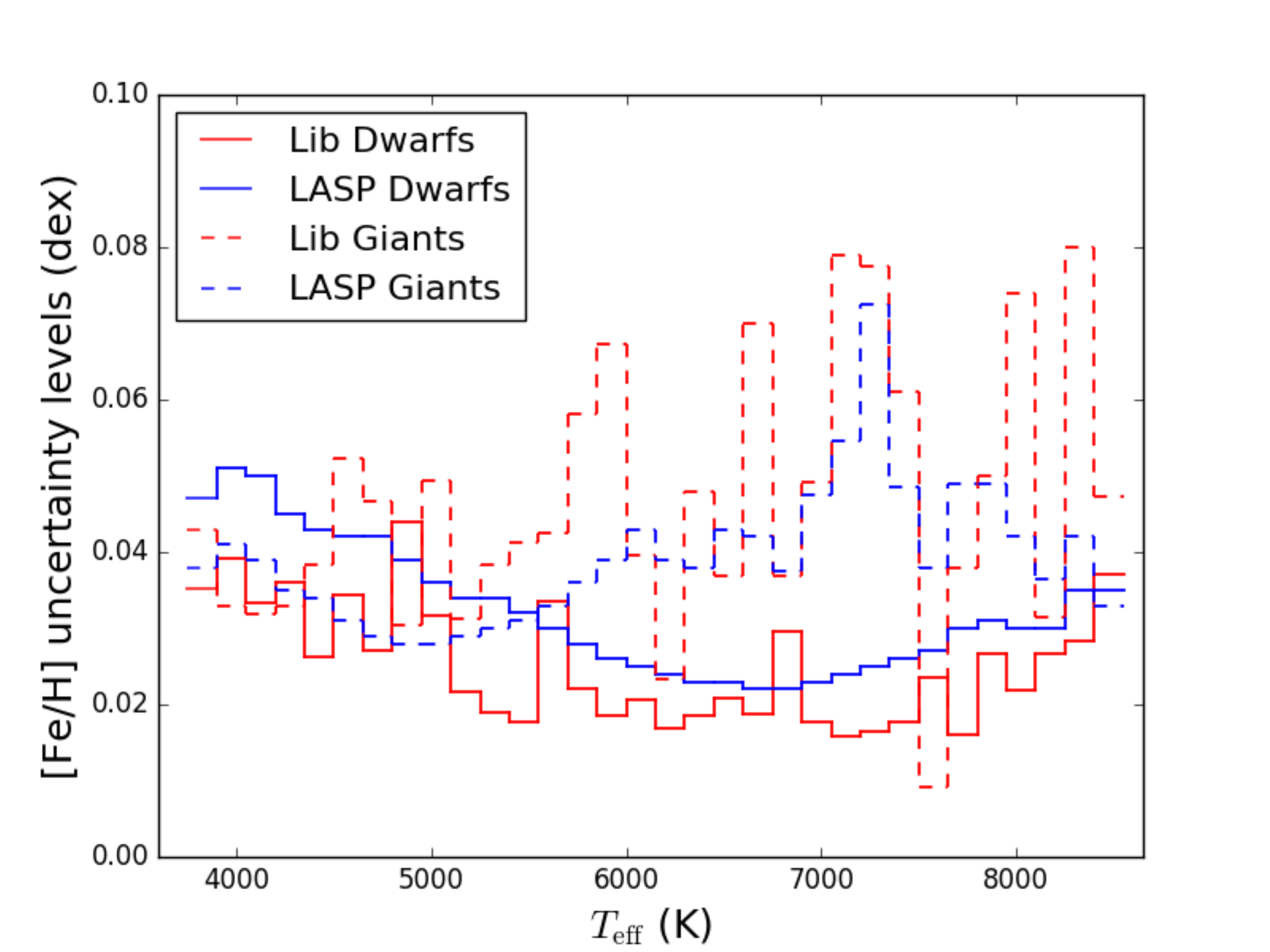}}
\end{minipage}
\begin{minipage}{0.3\linewidth}
  \centerline{\includegraphics[width=6cm]{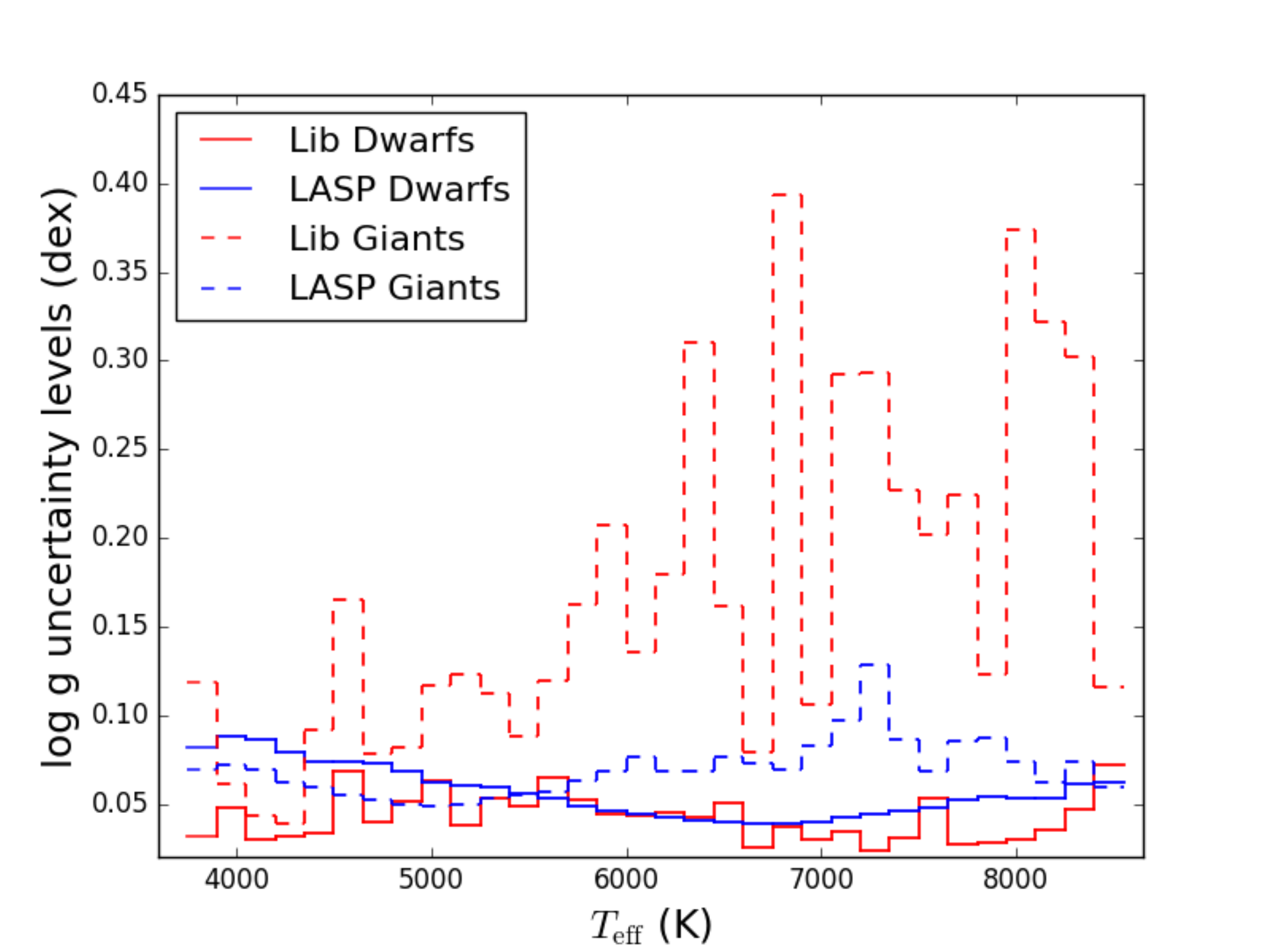}}
\end{minipage}
\caption{Distributions of the uncertainty levels of this library and LASP in the $T_{\rm eff}$  bins in steps of 150 K, left: $T_{\rm eff}$ uncertainty levels; middle: [Fe/H] uncertainty levels;  right:  log \emph{g} uncertainty levels. The blue (red) lines represent the results of LASP (our library), whereas the solid (dashed) lines represent the results of roughly dwarfs (giants).}
\label{fig19}
\end{figure}

 For a correct validation of the stellar parameter uncertainties of this library,  it is necessary to compare the labeled parameters for our templates to the reference parameters having  no or very small errors.  This would allow us to infer the precision from the standard deviation of the discrepancies and the accuracy from the systematic offsets. Unfortunately this measure is impossible because we have only the labeled parameters,  the reference parameters do not exist. If they were obtained from other sources, they would also be  affected by stochastic or systematic errors. Therefore, we analyzed only the uncertainty levels of parameters at different temperatures, which reflect the precision of parameters in regions of different temperatures. For this library, we compared the derived parameters from the internal  cross--validation to their library values. We separated the absolute values of differences into temperature bins in steps of 150 K.  For each  temperature bin  we took the median value of differences as its uncertainty level.  Moreover, we analyzed the parameter uncertainty levels at different temperatures of LASP. For the latter, we considered only spectra with high--SNRs (LAMOST SNRg $>$ 50.0). The uncertainties of their stellar parameters appeared to be primarily limited by the uncertainties in the library parameters. We separated the LASP--derived parameter errors into temperature bins in steps of 150 K.  For each temperature bin we took the median value of the errors as its  uncertainty level. We analyzed the  uncertainty levels  for giants (log \emph{g} $<$3.5) and roughly dwarfs (log \emph{g} $>$ 3.5),  separately. Fig \ref{fig19} shows the uncertainty levels at different temperatures. We found that the parameter uncertainty level  is closely related to the number of templates in the temperature range. With the increased  numbers of K giants in our library, their parameter uncertainty levels  are clearly lower than those of LASP. However, for other giants ( $T_{\rm eff } >$ 5500 K ), the limited number of observational spectra  used to co-add the template spectra resulted in a small number of templates in this parameter space.  Therefore, the parameter uncertainty levels of  hotter giants in this library is higher than those of LASP,  which added synthetic spectra in this parameter space.  For dwarfs, the parameter uncertainty levels of this library are lower than those of LASP in almost the entire temperature range except for a few individual grids.

Through comparisons with APOGEE and Kepler K giants, we confirmed that the systematic offsets of K giants between the spectroscopic and asteroseismic surface gravities were well--corrected for this library. However, a systematic overestimation of metallicities from this library, for [Fe/H] $<$ -1.5 dex was found  through comparisons with APOGEE and PASTEL.  Low metallicity stars were born in the early Universe and exists mainly in Galactic halo \citep{li}. The LAMOST experiment for Galactic Understanding and Exploration (LEGUE) spectroscopic survey of  the Galactic halo observed stars with r $<$ 16.8 mag at $|b| > 30^{\circ}$ \citep{lal}. Considering the time required to observe faint targets, they were observed only during dark nights. In addition, more stars were observed for the Galactic disk survey than for the  Galactic halo survey  though both covered the same area of the sky. Overall, more spectra of Galactic disk stars than Galactic halo stars were obtained by LAMOST.  Furthermore,  the spectral SNRs of low metallicity stars obtained by the LAMOST spectrograph were relatively low because most are distant stars.  Therefore, owing to the limited number (see Fig \ref{fig2}) and quality of the observational spectra for metal--poor stars, the co--added templates are very limited in this parameter space. With the development of more sophisticated telescopes, additional spectra of metal-poor stars will be available to create more complete empirical templates with spectra of higher quality.

\section{Conclusions}
\label{sect:conclusion}
Using data taken with the LAMOST spectrograph,  we built a library of low--resolution, high--quality optical spectra of 2892 touchstone stars with well--determined stellar parameters. The library with the standard deviations is available online in FITS formats\ \footnote{\url{http://paperdata.china-vo.org/empirical-lib/LAMOST-Empirical-lib/LAMOST-Emp-library.tar.gz}}.   The fundamental properties of a star can be extracted from its optical spectrum by comparison against this library. Our template library\\

\begin{enumerate}
\item covers the parameter space,  temperatures of 3750 K through 8500 K, metallicity from -2.5 dex to +1.0 dex, and log \emph{g} from 0 dex to 5.0 dex, with  grid steps of $\sim$150 K, 0.15 dex, and 0.25 dex for $T_{\rm eff}$, [Fe/H],  and log \emph{g},  respectively.
\item the wavelength coverage, 3800 \AA \ to 8900 \AA .
\item the spectral resolution, R $\sim$ 1800.
\end{enumerate}

Through internal cross--validation and external comparisons, we confirmed that the density of the library and  the quality of the associated stellar parameters enable the stellar parameter measurements  from this library to achieve  precisions of about 125 K in $T_{\rm eff}$, 0.1 dex in [Fe/H] and 0.20 dex in log \emph{g}, at a low spectral resolution R $\sim$ 1800.  Furthermore, the systematic offsets of K giants between the spectroscopic and asteroseismic surface gravities were well--corrected for our library.

A key advantage of this library is its completeness of empirical spectra with wavelength coverage of 3800 \AA -- 8900 \AA \  for K type stars,  particularly  cool stars of $T_{\rm eff}$ $<$ 4500 K.  By using empirical, rather than theoretical, spectra to build the library, we bypassed the difficulties that the current spectral synthesis codes have in modeling the complex spectra of cool stars. Moreover, the rich spectral information  contained in  the longer--wavelength spectra of these cool stars  is very important for many research topics related to cool stars.

\acknowledgments

The authors would like to thank Philippe Prugniel, Chao Liu, Hai--Ning Li and  Hai--Long, Yuan for helpful discussions. This work is supported  by the National Key Basic Research Program of China (Grant No.2014CB845700), National Science Foundation of China (Grant Nos 11703051, 11703053, and 11390371) and Key Research Program of Frontier Sciences, Chinese Academy of Sciences (Grant No. QYZDY-SSW-SLH007).
Guoshoujing Telescope (the Large Sky Area Multi-Object Fiber Spectroscopic Telescope, LAMOST) is a National Major Scientific Project built by the Chinese Academy of Sciences. Funding for the project has been provided by the National Development and Reform Commission. LAMOST is operated and managed by the National Astronomical Observatories, Chinese Academy of Sciences.



\end{document}